\documentclass[11pt]{article}
\usepackage{graphicx}
\usepackage{epsfig}
\usepackage{array}
\usepackage[figuresright]{rotating}
\usepackage{amsmath}
\usepackage{amssymb}
\usepackage{amsfonts}
\newcommand{\fpartial}{\mbox{$/\!\!\!\partial$}}

\newcommand{\pibf}{\mbox{\boldmath $\pi$}}
\newcommand{\taubf}{\mbox{\boldmath $\tau$}}

\newcommand{\veceps}{\mbox{\boldmath$\epsilon$}}
\newcommand{\vecsig}{\mbox{\boldmath$\sigma$}}
\newcommand{\vectau}{\mbox{\boldmath$\tau$}}

\renewcommand{\thefootnote}{\fnsymbol{footnote}}
\usepackage{cite}
\begin{document}

%%%%%%%%%%%%%%%%%%%%%%%%%%%%%%%%%%%%%%%%%%%%%%%%%%%%%%%%%%%%%%%%%%%%%%%%
%%%%%%%%%%%%%%%%%%%%%%% TexFileKennung und Datum %%%%%%%%%%%%%%%%%%%%%%%
%\begin{flushright}
%\footnotesize{up2p16/GENERAL/SCADRION/long-3.tex \today}
%\end{flushright}
%%%%%%%%%%%%%%%%%%%%%%%%%%%%%%%%%%%%%%%%%%%%%%%%%%%%%%%%%%%%%%%%%%%%%%%%

\begin{center}

{\LARGE{\bf Dispersion theory  of nucleon Compton scattering and
  polarizabilities}}\\[1ex] 
Martin Schumacher\footnote{Electronic address: mschuma3@gwdg.de}\\
II.  Physikalisches Institut der Universit\"at G\"ottingen,
Friedrich-Hund-Platz 1\\ D-37077 G\"ottingen, Germany\\
%Anatoly I. L'vov\footnote{Electronic address: lvov@x4u.lebedev.ru}\\
%P. N. Lebedev Physical Institute, RU-117924, Russia\\
Michael D. Scadron\footnote{Electronic address: scadron@physics.arizona.edu}\\
Physics Department, University of Arizona, Tucson, Arizona 85721, USA
\end{center}
\begin{abstract}
A status report on the topic  nucleon Compton scattering and
  polarizabilities is presented with emphasis on the scalar $t$-channel as
entering into dispersion theory. Precise values for the polarizabilities
  are obtained leading to
$\alpha_p=12.0\pm 0.6$ ($12.0$), $\beta_p=1.9\mp 0.6$ ($1.9$),
$\alpha_n=12.5 \pm 1.7$ ($13.4$), $\beta_n=2.7 \mp 1.8$ ($1.8$) 
in units of $10^{-4}$ fm$^3$
and 
$\gamma^{(p)}_\pi=-36.4\pm 1.5$ ($-36.6$), $\gamma^{(n)}_\pi= 58.6 \pm 4.0$ 
($58.3$), ($\gamma^{(p)}_0=-0.58\pm 0.20$), ($\gamma^{(n)}_0=+0.38\pm 0.22$)
in units of  
$10^{-4}$ fm$^4$, for the proton $(p)$ and neutron $(n)$, respectively. 
The data given with an error are $recommended$ experimental values with
updates compared to \cite{schumacher05} where necessary, 
the data in parentheses are predicted values. 
These predicted values are not contained in \cite{schumacher05}, but are the
  result of a newly developed analysis which is the main topic of the present
  paper. 
The most important recent discovery is that the largest part of the electric
polarizability and 
the total diamagnetic polarizability of the nucleon
are properties of the $\sigma$ meson as  part of the 
constituent-quark structure, as expected from the mechanism of chiral symmetry
  breaking. This view is supported by an experiment on Compton scattering by
  the proton carried out  in the second resonance region, where a
large contribution  from  the $\sigma$ meson enters into the scattering
amplitudes. This experiment led to a determination of the mass of the $\sigma$
meson of  $m_\sigma=600\pm 70$ MeV.
From the experimental $\alpha_p$ and predicted differences
  $(\alpha_n-\alpha_p)$ neutron polarizabilities in the range $\alpha_n =12.0 -
  13.4 $ are predicted, where the uncertainties are related to the
$f_0(980)$ and $a_0(980)$ scalar mesons.
\end{abstract}
\setcounter{footnote}{0}
\renewcommand{\thefootnote}{\arabic{footnote}}

\section{Introduction}

The idea to apply the coherent elastic scattering of photons (Compton
scattering ) to an investigation of the internal structure of the nucleon
dates back to the early 1950s. At that time  the structure of the nucleon
was mainly discussed in terms of a bare nucleon surrounded by a ``pion
cloud''.  In the framework of this simple model the polarization of the
pion  cloud  leads to an absorption of the incoming photon  and the emision
of the outgoing photon and thus to a Compton scattering process. 
Estimates carried out by Sachs and Foldy in 1950 showed that the pion-cloud 
polarization should lead to a small
modification of the Thomson scattering cross section
\cite{sachs50}. Later on in 1957 the  proposal was made by Aleksandrov 
\cite{aleksandrov57}
to measure  cross sections for the scattering of slow neutrons in the Coulomb
field of heavy nuclei. In this case it was predicted that the pion-cloud 
polarization leads to a small modification of Schwinger's scattering cross
section which takes into account the effects of the anomalous magnetic moment
of the neutron. For the proton the scattering in a Coulomb field is no
option because of the electric charge of the proton.

The first experiment on Compton scattering by the proton to measure
the polarizability was carried out in 1958. The final experiment of high
precision was performed at MAMI (Mainz) by Olmos de Le\'on et al in 2001
\cite{olmos01}. An extensive description of these experiments and their
analyses may be found in \cite{schumacher05,wissmann04,baranov01,drechsel03}  
The weighted averages of these results are given in
Table 1 of the present work.

First experiments on electromagnetic scattering of slow neutrons have been
carried out from 1986 to 1988. The results of these experiments 
are listed in Table 2 of 
\cite{schumacher05}. These  first experiments remained of rather limited 
precision. Great progress has been made in an experiment carried 
out by Schmiedmayer et al. in 1991 \cite{schmiedmayer91}.  The result 
obtained in \cite{schmiedmayer91} is
discussd in subsection 2.6 of the present paper.

As a second option to measure 
the electromagnetic polarizabilities of the neutron  quasifree 
Compton scattering by a neutron bound in a deuteron has been proposed by
Levchuk et al. \cite{levchuk94}. In this case only the internal 
structure of the neutron
is excited whereas the proton merely serves as a spectator. The first
high-precision experiment of this kind has been carried out by 
Kossert et al. in 2002 \cite{kossert02},
leading to a result which is discussed in subsection 2.6 of the present paper
together with the result of
electromagnetic scattering of neutrons \cite{schmiedmayer91}.

In principle also the cohenrent elastic Compton scattering by the deuteron
may be exploited. In this case  the sum of the proton and neutron 
polarizabilities is measured which may be used to arrive at the polarizability
of the neutron by subtracting the proton polarizability. A first experiment
of this kind has ben carried out by Lundin et al. \cite{lundin03}
and it has been proposed to repeat the experiment in order to arrive at a
higher precision \cite{feldman08}

\subsection{Theoretical approaches to the polarizabilities of the nucleon}

A number of theoretical papers concerned with the polarizabilities of the
nucleon based on nucleon models have been published in the 1980s. These
investigations  include the MIT bag model, the nonrelativistic quark model, 
the chiral quark model, the chiral soliton model and the Skyrme model.
The investigations  based nucleon models ceased with the advent of effective 
field theories at the beginning of the 1990s. These latter theories
attracted many researchers and  still are a subject of active investigations. 
Some information on the early activities may 
 be found in \cite{lvov93a,drechsel03,schumacher05}.
For the status of the present activities 
we wish to cite the 
recent article ``Using  effective field theory  to analysis  low-energy
Compton scattering data from protons and light nuclei''
\cite{griesshammer12} and ``Signatures of chiral dynamics in low-energy
Compton
scattering off the nucleon'' \cite{hildebrandt04}. In the analyses
described in \cite{griesshammer12}
the  polarizabilities are treated as free  parameters. The results obtained
may be compared with the present {\it recommended} values. 
In \cite{hildebrandt04} a projection formalism is presented which allows to
define dynamical polarizabilities of the nucleon from a multipole expansion of
the nucleon Compton amplitudes. These dynamical polarizabilities may be
compared with the corresponding quantities presented in section 6 of the
present work. Since these investigations 
\cite{griesshammer12,hildebrandt04}
have been published recently, 
the papers cited therein should give a complete overview.

\subsection{Outline of the present article}
The purpose of the present article is to give a concise 
status report on the research
topic of nucleon Compton scattering and polarizabilities, by describing
the results achieved on the basis of dispersion theory
since the latest review article published in 2005 
\cite{schumacher05}, partly based on published and partly on recent 
unpublished investigations. The published results are distributed over a 
larger number of papers
\cite{schumacher06,schumacher07a,schumacher07b,schumacher08,schumacher09,schumacher10,schumacher11a,schumacher11b}.
This makes it difficult for the reader to get insight into the present status
of development and makes a status report on the results highly advisable.

The present status report includes a complete precise prediction of
the polarizabilities and their interpretation in terms of degrees of freedom
of the nucleon. The progress made for the prediction of the 
nucleon structure ($s$-channel) part 
has been made possible by high-precision analyses \cite{drechsel07}
of  photomeson data and their conversion into partial
photoabsorption cross sections for all resonant and nonresonant
photoexcitation processes of the nucleon \cite{schumacher09,schumacher11a}.
These high-precision analyses contain the most precise information on the
electromagnetic structure of the nucleon available at present. Using
dispersion relations these partial photoabsorption cross sections can be
transferred into polarizabilities.  As a consequence we have 
a method to precisely predict the polarizabilities of the nucleon
on the basis of meson photoproduction data which includes all processes which
are  considered as the degrees of freedom of the nucleon.
However, when carrying out these predictions of electromagnetic
polarizabilities one is led to a severe problem. The electric polarizability
$\alpha$ predicted in this way amounts to only about 40\% of the experimental 
value and the magnetic polarizability $\beta$ overshoots the experimental
value by about a factor of 5. Apparently there is a missing piece in the 
electric polarizability and  a strong diamagnetic component which compensates
the predicted paramagnetic polarizability.

Predictions  based on models of the nucleon are in general not sensitive 
to details of the problems outlined above.
On the other hand, already in 1974 T.E.O. Ericson and coworkers
(BEFT) \cite{bernabeu74,bernabeu77} carried out calculations on the basis of
dispersion 
relations which  led to the discovery of the missing piece in the electric 
polarizability and to the missing diamagnetic polarizability. These authors
found a way of 
solving this severe problem by tracing it back to
the scalar $t$-channel contribution introduced by Hearn and Leader
in 1962 \cite{hearn62,koeberle68}. Quantitatively, the results obtained in
\cite{bernabeu74,bernabeu77} remained of 
rather limited precision. Therefore, in the following years further 
calculations were carried out by a number of researchers
with more or less success as far as the numerical results are
concerned. The main problem persisting up to the present  was to get data
of sufficient precision for the scalar two-photon excitation process
leading to a pion pair in the final  state.
This line of development is described in detail
in the previous review \cite{schumacher05}.

In \cite{schumacher05}
a step further is made by asking the question how this scalar $t$-channel
can be understood in terms of degrees of freedom of the nucleon, a question
which had never been asked before. Since
the structure of the nucleon as observed in photomeson experiments is
exhausted by the $s$-channel it was considered as likely that the
$t$-channel is related to the mesonic structure  of the constituent quarks.
This idea was worked out in a series of investigations 
\cite{schumacher06,schumacher07a,schumacher07b,schumacher08,schumacher09,schumacher10,schumacher11a,schumacher11b}
and proved to be extremely successful after it had been noted that
the quark-level linear $\sigma$ model (QLL$\sigma$M) as developed by
Delbourgo, Scadron and coworkers \cite{delbourgo95,beveren02,beveren09}
leads to a quantitative agreement with the experimental data. In this way
the missing part of the electric polarizability and the diamagnetic
polarizability were traced back to the $\sigma$ meson as part of the
constituent-quark structure.  From the point of view of chiral symmetry
breaking this does not come as a surprise. Chiral symmetry breaking explains
the mass of the constituent quark in terms of the exchange of the $\sigma$
meson with $q\bar{q}$ loops of the QCD vacuum. This necessarily leads 
to a $\sigma$ meson as part of the constituent-quark structure.

The nucleon model introduced above containing a $\sigma$ meson as part of the
constituent-quark structure is strongly supported by a Compton scattering
experiment on the proton  carried out at
MAMI (Mainz) in the 1990s and published in 2001 \cite{galler01,wolf01}. 
In this experiment it has been shown that
the scalar $t$-channel 
makes a strong contribution to the Compton scattering 
amplitude in the second resonance region of the nucleon at large scattering
angles.  This contribution  is  successfully  represented in terms of a 
$t$-channel pole located at $m^2_\sigma$ as introduced by L`vov et al.
in 1997 \cite{lvov97}
where  $m_\sigma$ is 
the ``bare'' mass\footnote{Strictly speaking the bare mass is not a measurable
quantity. The term ``bare'' mass introduced here is used to denote a particle
with no open decay channel except for $\gamma\gamma$ (see subsection 5.4).} 
of the $\sigma$ meson, determined in this experiment to be 
$m_\sigma= 600\pm 70$ MeV. The error given here corresponds to the
uncertainty of the CGLN amplitude representing the $P_{33}(1232)$ resonance. 
Details are given in section 6.
In spite of this great success the physical interpretation of the experiment 
remained uncertain because the  $\sigma$ meson was not a generally accepted
particle at that time. This led to a delay in the final interpretation
of the experiment, because  additional detailed theoretical investigations 
were required 
in oder to remove the remaining uncertainties.
The breakthrough came in 2006
\cite{schumacher06} when it was shown that it is possible to
quantitatively predict the scalar
$t$-channel part of the electromagnetic polarizabilities
in terms of the QLL$\sigma$M and thus
proving that the $t$-channel part of the Compton-scattering amplitude
is a property of the structure of 
the constituent quarks. This result was further worked out in
a series of additional investigation \cite{schumacher07a,schumacher07b,schumacher08,schumacher09,schumacher10,schumacher11a}.

In the course of these further investigations it was realized that the $\sigma$
meson provides by far   the largest part of the scalar $t$-channel
contribution. But there are also smaller contributions from the $f_0(980)$
and $a_0(980)$ scalar mesons. These mesons together with the $\sigma$ meson
are members of the scalar nonet $\sigma(600)$, $\kappa(800)$, $f_0(980)$
$a_0(980)$. This observation led to the attempt to extend the 
QLL$\sigma$M of 
\cite{delbourgo95,beveren02,beveren09}
 to SU(3). The investigation carried out in \cite{schumacher11b} 
showed that the members of the 
scalar nonet  have the same mass
in the chiral limit (cl) amounting  to $m^{\rm cl}_\sigma=652$ MeV. This 
degeneracy is removed via explicit symmetry breaking by taking into account
the effects of the finite current-quark masses, applying the rules of 
spontaneous and dynamical symmetry breaking.

The present review is organized as follows. In section 2 we present an updated
list of {\it recommended} experimental values for the polarizabilities. New
arguments are given in favor of the value 
$\alpha_n=(12.5\pm 1.7)\times 10^{-4}$fm$^3$ of the neutron electric
polarizability. A slight modification
of the {\it recommended} value of $\gamma^{(p)}_\pi$ is found.
In section 3 we derive a complete list of predicted contributions to the
$s$-channel polarizabilities. In section 4 we describe the properties of the
invariant amplitudes and their relation to the polarizabilities.
Section 5 is concerned with the $t$-channel polarizabilities. Since the methods
and results for the $t$-channel are widely unknown we provide an extended
description of the $t$-channel aimed to arrive at high clarity. In section 6
we describe Compton scattering in a large angular range and at energies up to
1 GeV, with emphasis on interference effects between nucleon structure
($s$-channel) and constituent-quark structure ($t$-channel) contributions.

\section{Definition of polarizabilities and compilation of experimental
results}

In the following we will outline the basic relations  used
in our studies of  nucleon Compton scattering and polarizabilities
beginning with a classic
approach to the polarizabilities. We will see that in addition
to the electromagnetic polarizabilities $\alpha$ and $\beta$
spin-polarizabilities  $\gamma_0$ and $\gamma_\pi$
for Compton scattering in the forward and backward directions are required.
The experiments and  the experimental results have been described in 
previous works \cite{wissmann04,schumacher05}. This makes it sufficient for the present review
to only summarize the experimental
results

\subsection{Polarizabilities and  electromagnetic fields \label{2.1}}

A nucleon in an electric field ${\bf E}$ and  magnetic field ${\bf H}$ obtains
an induced electric dipole moment ${\bf d}$ and magnetic dipole 
moment ${\bf m}$
given by 
\begin{eqnarray}
&&{\bf d}= 4\pi \,\alpha \,{\bf E}, \label{electric}\\
&&{\bf m}= 4\pi\, \beta \,{\bf H}, \label{magnetic}
\end{eqnarray}
in a unit system where the electric charge $e$ is given by  
$e^2/4\pi =\alpha_{em}=1/137.04$. Eqs. (\ref{electric})
and (\ref{magnetic}) may be understood as the response of the nucleon 
to the fields provided by a real or virtual photon and it is evident
that we need a second photon to measure the polarizabilities. This may be
expressed through the relation
\begin{equation}
H^{(2)}=-\frac12\,4\pi\,\alpha\,{\bf E}^2 -\frac12\,4\pi\,\beta\,{\bf H}^2
\label{energy}
\end{equation}
where $H^{(2)}$ is the energy change in the electromagnetic field due to the
presence of the nucleon in the field.

\subsection{Forward and backward Compton scattering \label{2.2}}

Compton scattering contains  the most effective method to provide two photons
interacting simultaneously with the nucleon. Therefore, Compton scattering is
the method of first choice for the measurement of polarizabilities. The
second method is the electromagnetic scattering of slow neutrons in the
electric field of a heavy nucleus.

The differential cross section for Compton scattering
\begin{equation}
\gamma N \to \gamma' N'
\end{equation}
may be written in the form \cite{babusci98}
\begin{equation}
\frac{d \sigma}{d \Omega}= \Phi^2 |T_{fi}|^2
\end{equation}
with $\Phi=\frac{1}{8 \pi M}\frac{\omega'}{\omega}$ in the lab frame
($\omega$, $\omega'$ = lab energies of the incoming and outgoing photon)
and $\Phi=\frac{1}{8 \pi \sqrt{s}}$ in the cm frame 
($\sqrt{s}$ = total energy). For the
following discussion it is convenient to use the lab frame and to
consider special cases for the amplitude $T_{fi}$.
These special cases are
the extreme forward ($\theta=0$) and extreme backward ($\theta=\pi$)
direction  where the amplitudes for Compton scattering
may be written in the form
\cite{babusci98}  
\begin{eqnarray}
&&\frac{1}{8\pi m}[T_{fi}]_{\theta=0}
=f_0(\omega){\veceps}'{}^*\cdot{\veceps}+
g_0(\omega)\, \mbox{i}\, {\vecsig}\cdot({\veceps}'{}^*\times
{\veceps}), \label{T0}\\
&&\frac{1}{8\pi m}[T_{fi}]_{\theta=\pi}
=f_\pi(\omega){\veceps}'{}^*\cdot{\veceps}+
g_\pi(\omega)\, \mbox{i}\,{\vecsig}\cdot({{\veceps}}'{}^*
\times
{{\veceps}}), 
\label{Tpi}
\end{eqnarray}
with $m$ being the nucleon mass, ${\veceps}$ the polarization of the photon
and $\vecsig$ the spin vector. 

The process described in (\ref{T0}) is the
transmission of linearly polarized photons through a medium as
provided by a nucleon with spin vector $\vecsig$ parallel or
antiparallel to the direction of the incident photon
with rotation of the direction of linear
polarization. The amplitudes   $f_0$ and $g_0$   correspond
to the  polarization components of the outgoing photon 
parallel and perpendicular to the direction of linear polarization
of the incoming photon. The
interpretation of (\ref{Tpi}) is the same as that of (\ref{T0})
except for the fact that the photon is reflected. In case of forward
scattering  (\ref{T0})   the origin of 
the rotation of the  direction of linear polarization may be
related partly to the alignment of the  internal magnetic dipole moments of
the  nucleon and partly to the dynamics of the pion cloud.  In case of backward
scattering  (\ref{Tpi}) a large portion of the relevant phenomenon
is of a completely different origin. This is  the $t$-channel contribution 
which will be studied in detail 
in the following.
It is important to realize that in contrast to 
frequent belief  
there is no flip of any spin in the two cases of Compton scattering.
The factor $\vecsig$ in the second terms of the two equations may be
interpreted as a spin dependence of scattering in the sense that the
direction of rotation of the electric vector changes sign (from $e.g.$
clock-wise to anti clock-wise) when the spin of the target nucleon
is reversed. 

The two scattering processes may also be related to the two states of
circular polarization, i.e. helicity amplitudes for forward and
backward Compton scattering. 
If the photon and nucleon spins are parallel
(photon helicity $\lambda_\gamma=+1$, nucleon helicity 
$\lambda_N= -\frac12$, and net helicity
in the photon direction 
$\lambda =\lambda_\gamma-\lambda_N= \frac32$) 
then the amplitude is
\begin{equation}
f^{3/2}_0(\omega)=f_0(\omega)-g_0(\omega), \label{f32}
\end{equation}
while for  the spins being  anti-parallel (photon helicity 
$\lambda_\gamma=+1$, nucleon
helicity $\lambda_N=+\frac12$, and net helicity along the photon direction
$\lambda=\lambda_\gamma-\lambda_N=+\frac12$) the amplitude is
\begin{equation}
f^{1/2}_0(\omega)=f_0(\omega)+g_0(\omega). \label{f12}
\end{equation}
The amplitudes $f^{3/2}_0$ and $f^{1/2}_0$ are related by the optical
theorem to the total cross sections $\sigma_{3/2}$ and $\sigma_{1/2}$
for the reaction 
\begin{equation}
\gamma+N \to N^* \to N + {\rm mesons} + \mbox{radiative decay of N$^*$}
\label{absorption}
\end{equation}
when the photon spin is parallel or anti-parallel to the nucleon spin:
\begin{eqnarray}
&&{\rm Im}f^{3/2}_0(\omega) =\frac{\omega}{4\pi}\sigma_{3/2}(\omega),\\
&&{\rm Im}f^{1/2}_0(\omega) =\frac{\omega}{4\pi}\sigma_{1/2}(\omega).
\label{optical-1}
\end{eqnarray}
Therefore,
\begin{eqnarray}
&&{\rm Im}\, f_0(\omega)
=\frac{\omega}{4\pi}\, \frac{\sigma_{1/2}(\omega)+\sigma_{3/2}(\omega)}{2}
=\frac{\omega}{4\pi}\sigma_{\rm tot}(\omega),\label{optical-3}\\
&&{\rm Im}\, g_0(\omega)=\, \frac{\omega}{4\pi}\frac{\sigma_{1/2}(\omega)
-\sigma_{3/2}(\omega)}{2}=\frac{\omega}{8\pi}\,\Delta\sigma(\omega),
\label{optical-4}
\end{eqnarray}
where $\sigma_{\rm tot}(\omega)=(\sigma_{1/2}+\sigma_{3/2})/2$
is the spin averaged total cross section and $\Delta\sigma=
\sigma_{1/2}-\sigma_{3/2}$.
For the backward direction we obtain
\begin{eqnarray}
{\rm Im}f_\pi(\omega)&=&\frac{\omega}{4\pi}\left[\sigma(\omega,
\Delta P={\rm yes}) - \sigma(\omega,\Delta P={\rm no})\right],
\label{optical-5}\\
{\rm Im}g_\pi(\omega)&=&\frac{\omega}{8\pi}\left[\Delta\sigma(\omega,
\Delta P={\rm yes}) - \Delta\sigma(\omega,\Delta P={\rm no})\right],
\nonumber\\
&\equiv&\frac{\omega}{8\pi}\{ [\sigma_{1/2}(\omega,\Delta P={\rm
yes})-  \sigma_{1/2}(\omega,\Delta P={\rm no}) ],\nonumber\\&&\,\,\,\,\,\,-
[\sigma_{3/2}(\omega,\Delta P={\rm
yes})-  \sigma_{3/2}(\omega,\Delta P={\rm no}) ],\}\nonumber\\
&\equiv& \frac{\omega}{8\pi}\sum_n P_n[\sigma^n_{3/2}(\omega)
-\sigma^n_{1/2}(\omega)]\label{optical-7}
\end{eqnarray}
with $P_n=+1$ for $\Delta P=$ no (no parity change) and  $P_n=-1$ for 
$\Delta P=$ yes (with parity change).  
Details of the derivation may be found in \cite{schumacher05}.

\subsection{Definition of polarizabilities}

Following Babusci et al. \cite{babusci98} the equations (\ref{T0})
and (\ref{Tpi}) can be used to define the electromagnetic
polarizabilities and spin-polarizabilities as the lowest-order
coefficients in an $\omega$-dependent development of the
nucleon-structure dependent parts of the scattering amplitudes:
 \begin{eqnarray}
f_0(\omega) & = & - ({e^2}/{4 \pi m})Z^2 + 
{\omega}^2 ({\alpha}
+{\beta}) + {\cal O}({\omega}^4), \label{f0}\\
g_0( \omega) &=&  \omega\left[ - ({e^2}/{8 \pi m^2})\,   
{\kappa}^2 
+ {\omega}^2
{\gamma_0}  + {\cal O}({\omega}^4) \right], \label{g0}\\
f_\pi(\omega) &=& \left(1+({\omega'\omega}/{m^2})\right)^{1/2} 
[-({e^2}/{4 \pi m})Z^2 +       
\omega\omega'({\alpha} - {\beta}) 
+{\cal O}({\omega}^2{{\omega}'}^2)], \label{fpi}\\
g_\pi(\omega) &=& \sqrt{\omega\omega'}[
({e^2}/{8 \pi m^2})  
( {\kappa}^2 + 4Z
{\kappa} + 2Z^2)
+ \omega\omega'
{{\gamma_\pi}} + {\cal O}
({\omega}^2 {{\omega}'}^2)],  \label{gpi}
\end{eqnarray}
where $Z e$ is the electric charge ($e^2/4\pi=1/137.04$), 
$\kappa$ the anomalous magnetic
moment of the nucleon and $\omega'=\omega/(1+\frac{2\omega}{m})$.

In the relations for $f_0(\omega)$ and $f_\pi(\omega)$ the first
nucleon structure dependent coefficients are the photon-helicity non-flip 
$(\alpha+\beta)$ and photon-helicity flip $(\alpha-\beta)$ linear
combinations of the electromagnetic polarizabilities $\alpha$ and
$\beta$. In the relations for $g_0(\omega)$ and $g_\pi(\omega)$
the corresponding coefficients are the spin polarizabilities
$\gamma_0$ and $\gamma_\pi$, respectively. 

\subsection{Interpretation of Compton scattering in terms of electric and
  magnetic fields \label{2.4}}

\begin{figure}[h]
\begin{center}
\includegraphics[width=0.5\linewidth]{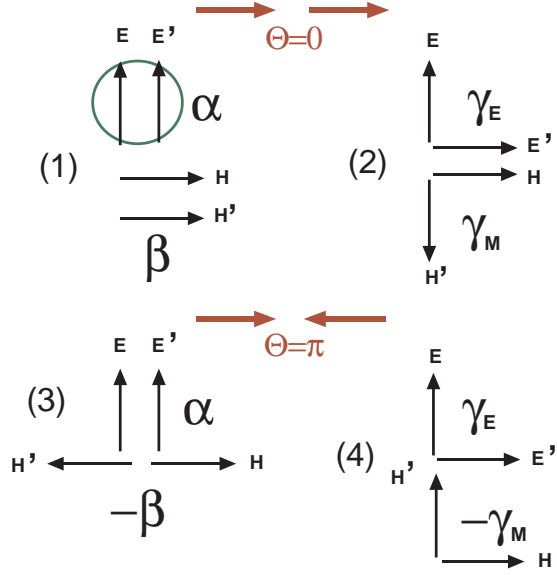}
\end{center}
\caption{Compton scattering viewed as simultaneous interaction of two electric
field vectors ${\bf E,E'}$ and magnetic field vectors ${\bf H,H'}$ for four
different cases. (1) Helicity independent forward Compton scattering as given
by the amplitude $f_0(\omega)$. (2) Helicity dependent forward Compton 
scattering as given by the amplitude $g_0(\omega)$. (3) Helicity independent
backward Compton scattering as given by the amplitude $f_\pi(\omega)$.
(4) Helicity dependent backward Compton scattering as given by the amplitude
$g_\pi(\omega)$. Longitudinal photons can only provide two electric vectors
with parallel planes of linear polarization as shown in the encircled  upper
part of panel (1). In panels (2) and (4) the direction of rotation leading
from ${\bf E}$ to ${\bf E'}$ depends on the helicity difference
$|\lambda_p-\lambda_\gamma|$. }
\label{ehfields}
\end{figure}

Polarizabilities may be measured by simultaneous interaction of two photons
with the nucleon. In case of static fields this may be written in the form
given by Eq. (\ref{energy}). The first part on the r.h.s. of Eq. (\ref{energy})
is realized in experiments where slow neutrons are scattered in the electric
field of a heavy nucleus, leading to a measurement of the electric
polarizability as described by Eq. (\ref{neutronscatt}). The scattering
of slow neutrons in the electrostatic field of a heavy nucleus corresponds to
the encircled  upper part of panel (1) in Figure 1, showing two parallel
electric vectors. Only this case is accessible with longitudinal photons
at low particle velocities.
Compton scattering in the forward and backward directions leads
to more general combinations of electric and magnetic fields. 
This is depicted in the four panels of  Figure \ref{ehfields}.
Panel (1) corresponds to the amplitude $f_0(\omega)$, panel (2) to the
amplitude $g_0(\omega)$, panel (3) to the amplitude $f_\pi(\omega)$ and
panel (4) to the amplitude $g_\pi(\omega)$. Panel (1) contains two parallel
electric vectors and two parallel magnetic vectors. Through these
the electric polarizability $\alpha$ and the magnetic polarizability $\beta$ 
can be measured. In panel (3) corresponding to backward scattering the
direction of the magnetic vector is reversed so that instead of $+\beta$
the quantity $-\beta$ is measured. The nucleon  has a spin and because of 
this the
two electric and magnetic vectors have the option of being perpendicular
to each other.  This leads to the definition of the spin polarizability
$\gamma$ which comes in different versions $\gamma_E$ and $\gamma_M$,
respectively. The different directions of the magnetic field in the forward
and backward direction leads to definitions of polarizabilities for the
four cases
\begin{eqnarray}
&&(1)\quad (\alpha+\beta)\hspace{15mm}\text{(forward polarizability)},
\label{eq11}\\
&&(2) \quad \gamma_0= \gamma_E+\gamma_M \hspace{5mm}
\text{(forward spin-polarizability)}, \label{eq12}\\
&&(3)\quad (\alpha-\beta) \hspace{15mm}\text{(backward polarizability)},
\label{eq13}\\
&&(4) \quad \gamma_\pi=(\gamma_E - \gamma_M) \hspace{2mm}
\text{(backward  spin-polarizability)}.
\label{eq14}
\end{eqnarray}

In former descriptions of Compton scattering and polarizabilities
occasionally ``Compton polarizabilities'' $\overline{\alpha}$ 
and $\overline{\beta}$
have been introduced in order to distinguish them from the ``static''
polarizabilities $\alpha$ and $\beta$. The advantage of Figure  
\ref{ehfields} is that firm arguments are presented that a distinction
of this kind is quite unnecessary.

\subsection{Sum rules and $s$-channel polarizabilities}

At this point it is
useful to give expressions for sum rules which will be used later for the
prediction of
 the polarizabilities. The following dispersion integrals have been derived:
\begin{eqnarray}
&&\alpha+\beta=\frac{1}{2\pi^2}\int^\infty_{\omega_0}
\frac{\sigma_{tot}(\omega)}{\omega^2}d\omega, \label{baldin}\\
&&\gamma_0=-\frac{1}{4\pi^2}\int^\infty_{\omega_0}
\frac{\sigma_{3/2}-\sigma_{1/2}}{\omega^3}d\omega,  \label{spinpol}\\
&&(\alpha-\beta)^s=\frac{1}{2\pi^2}\int^\infty_{\omega_0}\sqrt{1+\frac{2\omega}{m}} 
[\sigma(\omega,\Delta P={\rm yes})-\sigma(\omega,\Delta P={\rm no})]\frac{d\omega}{\omega^2},
\label{BEFT}\\
&&\gamma^s_\pi=\frac{1}{4\pi^2}\int^\infty_{\omega_0} \sqrt{1+\frac{2\omega}{m}}
\left(1+\frac{\omega}{m}\right)\sum_nP_n[\sigma^n_{3/2}(\omega)-
\sigma^n_{1/2}(\omega)]\label{LN}\frac{d\omega}{\omega^3}
\end{eqnarray}
where $P_n=+1$ for $\Delta P=$ no (no parity change) and $P_n=-1$ 
for $\Delta P=$ yes (with parity change).
The sum rule for $(\alpha+\beta)$ is the Baldin-Lapidus (BL) sum rule
\cite{baldin60},
the sum rule for $\gamma_0$ a subtracted version of the
Gerasimov-Drell-Hearn (GDH) sum rule \cite{gerasimov66}, 
the sum rule for $(\alpha-\beta)^s$
the $s$-channel part of the Bernabeu et al. (BEFT) sum rule 
\cite{bernabeu74,bernabeu77} and the sum rule
for $\gamma^s$ the $s$-channel part of the L'vov-Nathan (LN) sum rule
\cite{lvov99}.
The two latter sum rules have to be supplemented by $t$-channel pole terms as
shown in the following.

The appropriate tool for the prediction of  the polarizabilities 
$\alpha$ and $\beta$
is to simultaneously apply
the forward-angle sum rule for $(\alpha+\beta)$ and the backward-angle sum
rule for $(\alpha-\beta)$. This leads to the following relations
\cite{schumacher07a,schumacher09} : 
\begin{eqnarray}
&&\alpha=\alpha^s+\alpha^t, \label{polar1}\\
&&\alpha^s=\frac{1}{2\pi^2}\int^\infty_{\omega_0}\left[A(\omega)\sigma(\omega,E1,M2,\cdots)+B(\omega)\sigma(\omega,M1,E2,\cdots)\right]\frac{d\omega}{\omega^2},\label{polar2}\\
&&\alpha^t=\frac12(\alpha-\beta)^t
\end{eqnarray}
and
\begin{eqnarray}
&&\beta=\beta^s+\beta^t,\label{polar3}\\
&&\beta^s=\frac{1}{2\pi^2}\int^\infty_{\omega_0}\left[A(\omega)
\sigma(\omega,M1,E2,\cdots)+B(\omega)\sigma(\omega,E1,M2,\cdots)\right]
\frac{d\omega}{\omega^2},\label{polar4}\\
&&\beta^t=-\frac12(\alpha-\beta)^t,
\end{eqnarray}
with
\begin{eqnarray}
&&\omega_0=m_\pi+\frac{m^2_\pi}{2 m},\label{polar5}\\
&&A(\omega)=\frac12 \left( 1+\sqrt{1+\frac{2\omega}{m}} \right),
\label{poarl6}\\
&&B(\omega)=\frac12 \left( 1-\sqrt{1+\frac{2\omega}{m}} \right).
\label{polar7}
\end{eqnarray}
In (\ref{polar1}) to (\ref{polar7}) $\omega$ is the photon energy in the
lab system and  $m_\pi$
the pion mass.  The quantities $\alpha^s,\beta^s$
are the $s$-channel electric and magnetic polarizabilities, and 
$\alpha^t,\beta^t$ the $t$-channel electric and magnetic polarizabilities,
respectively. The multipole content of the photo-absorption cross-section
enters through
\begin{eqnarray}
&&\sigma(\omega,E1,M2,\cdots)=\sigma(\omega,E1)+ \sigma(\omega,M2)+\cdots,
\label{polar9}\\
&&\sigma(\omega,M1,E2,\cdots)=\sigma(\omega,M1)+ \sigma(\omega,E2)+\cdots,
\label{polar20}
\end{eqnarray}
{\rm  i.e.} through the sums of cross-sections with change and without 
change of parity during the electromagnetic transition, respectively. The
multipoles belonging to parity change are favored for the
electric polarizability $\alpha^s$ whereas the multipoles belonging to parity
nonchange are favored for the magnetic polarizability $\beta^s$. For the
$t$-channel parts we use the pole representations described in
\cite{schumacher07b} and the present section 5. 

It is interesting to note that the quantities $A(\omega)$ and $B(\omega)$
tend to 1 and 0, respectively, for   $ m \to \infty$. This implies that the
mixing of parity taking place in (\ref{polar2}) and (\ref{polar4}) vanishes 
for an infinite nucleon mass. This shows that the mixing of parity may be 
interpreted as a ``recoil'' effect.

\subsection{Experimental results obtained for the polarizabilities}

Static electric fields ${\bf E}$ with sufficient strength are provided by heavy
nuclei. Therefore, use may be made of the differential cross section 
for the electromagnetic scattering of slow neutrons in the Coulomb field
of heavy nuclei \cite{lvov93a}
\begin{equation}
\frac{d \sigma_{\rm pol}}{d\Omega}=\pi \, m \,p\,(Ze)^2\,{\rm Re}\,a
\left\{\alpha_n\sin\frac{\theta}{2}-\frac{e^2 \kappa^2_n}{2 m^3}
\left(1-\sin\frac{\theta}{2}\right)\right\}
\label{neutronscatt}
\end{equation}
where $p$ is the neutron momentum and $-a$ the amplitude for hadronic 
scattering by the nucleus, $m$ the neutron mass and $(Ze)$ the charge of the 
scattering nucleus. The second term in the braces is due to the
Schwinger term, i.e. the term describing neutron scattering in the Coulomb
field due to the magnetic moment of the neutron only.

Electromagnetic scattering of neutrons has been carried out in a larger number
of experiments where in the latest \cite{schmiedmayer91} the  number
\begin{equation}
\alpha_n=12.6\pm 2.5\,\,\,10^{-4}{\rm fm}^3\quad (\text{Coulomb scattering of
  neutrons}) 
\label{schmiedmayer}
\end{equation}
has been obtained after a correction has been carried out \cite{lvov93a}
taking into account  that in the original evaluation the second term
in curly brackets of (\ref{neutronscatt}) has been ignored. 
In \cite{lvov93a} the original result evaluated by \cite{schmiedmayer91}
$\alpha_{0n}=12.0\pm 1.5 \pm 2.0$ is corrected leading to 
$\alpha_n=12.6\pm 1.5 \pm 2.0$.
The corrected result of this
last experiment on electromagnetic scattering of neutrons is very reasonable,
where reasonable means that there is a good agreement with a
later experiment \cite{kossert02} where  quasifree Compton scattering
by a neutron, originally bound in the deuteron is exploited. In the latter
case 
\begin{equation}
\alpha_n=12.5\pm 2.3\,\,\,10^{-4}{\rm fm}^3\quad(\text{quasi-free Compton
  scattering by the neutron})
\label{kossert}
\end{equation}
is obtained. Furthermore, there are activities to exploit coherent Compton
scattering by the deuteron and the result of a first successful experiment 
has been published \cite{lundin03}. These experiments are continued
in order to improve on the precision \cite{feldman08}.
In addition to  these experiments on the neutron there is a large number of
Compton scattering experiments on the proton which have been described
in the
previous summaries \cite{wissmann04,schumacher05}.

From these data
{\it  recommended} polarizabilities are found and given in 
Table \ref{exresults}.
\begin{table}[h]
\caption{Summary of experimental results on the polarizabilities.
The BL sum-rule constraints are $\alpha_p+\beta_p= 13.9\pm 0.3$, 
$\alpha_n+ \beta_n= 15.2\pm 0.5$}
\vspace{3mm}
\begin{tabular}{l}
\hline
\\
\vspace{3mm}
$\alpha_p=12.0\pm 0.6$, $\beta_p=1.9\mp 0.6$,
$\alpha_n=12.5 \pm 1.7$, $\beta_n=2.7 \mp 1.8$ \,\,in units of $10^{-4}$
fm$^3$\\ \vspace{3mm}
$\gamma^{(p)}_\pi=-36.4\pm 1.5$, $\gamma^{(n)}= 58.6 \pm 4.0$ in units of 
$10^{-4}$ fm$^4$\\
\hline
\end{tabular}
\label{exresults}
\end{table}
The numbers given here are the same as those of \cite{schumacher05} except for
$\gamma^{(p)}_\pi$ where an update was necessary which was discussed in
\cite{schumacher11a} and the present subsection 6.5.

\subsection{Neutron polarizability $\alpha_n$ determined from the
  experimental $\alpha_p$ and the predicted difference $(\alpha_n-\alpha_p)$}

In addition to the  experimental results given in (\ref{schmiedmayer})
and (\ref{kossert}) we may take
into consideration a determination of the neutron polarizability 
from the proton polarizability by making use of predicted isovector components
of the  nucleon polarizabilities. This method has been
proposed and applied  for the first time in \cite{schumacher07b}.
The relevant numbers  may be found in \cite{schumacher07b} and in
Tables  \ref{polresults1} and \ref{tab4}    of the present work. As shown
before 
the electromagnetic polarizabilities consist of an $s$-channel contribution
due to the excited states of the nucleon and a $t$-channel contribution 
due to the scalar mesons $\sigma$, $f_0(980)$ and $a_0(980)$. For the
proton the present  method of calculation
leads to predictions which are in  excellent agreement
with the experimental electromagnetic polarizabilities.
Since the
experimental value for $\alpha_p$ has an error of only 5\% we may 
tentatively conclude that the present method of predicting the
electric polarizabilities of the proton and the neutron 
has been tested at a level of precision of 5\%. 
However, there is a caveat related to the 
$t$-channel
contribution stemming from the $f_0(980)$ and $a_0(980)$ mesons
which is predicted to be given by
\begin{equation} 
\alpha(f_0(980),a_0(980))=+0.3-0.4\,\tau_3.
\label{alphafoao}
\end{equation}
There are very good reasons that this prediction is correct with good
precision. But it would be of advantage to have an experimental confirmation.
The excellent agreement between 
predicted and experimental polarizabilities observed for the proton 
means that 
the $s$-channel prediction is confirmed to be precise, but for the   
$t$-channel 
only the $\sigma$-meson part and the 
cancellation of the $f_0(980)$ and $a_0(980)$ contributions are  confirmed
but not the sign of $\alpha(f_0(980),a_0(980))$. This sign can only be
determined by a high-precision experiment on the neutron and has to be left
as an open problem as long as such an experiment has not been carried out.
Because of this we give in the following predictions
for $\alpha_n$ including and not including the effects of $f_0(980)$ and
$a_0(980)$ and the latter also with a reversed sign. These predictions are
\begin{eqnarray}
&&\alpha_n=+13.4\,\, 10^{-4}{\rm fm}^3 \quad (\text{
  $f_0(980)$ and $a_0(980)$ included}),\\
&&\alpha_n=+12.7\,\, 10^{-4}{\rm fm}^3\quad (\text{
  $f_0(980)$ and $a_0(980)$ not included} ),\\
&&\alpha_n=+12.0\,\, 10^{-4}{\rm fm}^3\quad (\text{
  $f_0(980)$ and $a_0(980)$ included with reversed signs}).
\label{protonneutron}
\end{eqnarray}
Apparently this rather straightforward determination of the electric 
polarizability of the neutron  strongly supports the experimental
results given in (\ref{schmiedmayer}) and (\ref{kossert}). 
Furthermore, there is a tendency towards a larger  value for $\alpha_n$
rather than a smaller value than $12.0$ (see PDG 2012 \cite{PDG}). 
This  gives us
confidence that the weighted average of  the two experimental
results in  (\ref{schmiedmayer}) and (\ref{kossert}) is a good choice
for a $recommended$ final result as given in Table \ref{exresults}.
For the experiments to come it can be predicted that the neutron electric
polarizability obtained in a high-precision experiment should be 
close to the interval $\alpha_n=12.0 - 13.4$.

\section{Photoabsorption and $s$-channel polarizabilities}

In the following partial contributions to the $s$-channel part of the
polarizabilities are calculated from the corresponding partial photoabsorption
cross sections using dispersion relations. Therefore, we first have to explain
how these partial photoabsorption cross sections are generated. Two different
approaches are possible which will be described in the following. The first
makes use of total photoabsorption cross sections analyzed in terms of
resonant and nonresonant partial cross sections \cite{armstrong72} and 
the other of high-precision analyses \cite{drechsel07 }
of photomeson data and their
conversion into partial photoabsorption cross sections.

\subsection{The total photoabsorption cross section}

\begin{figure}[h]
\begin{center}
\includegraphics[width=0.45\linewidth]{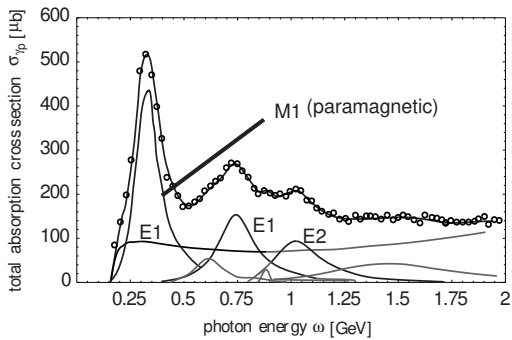}
\end{center}
\caption{Total photoabsorption cross section decomposed 
into resonant and nonresonant components \cite{armstrong72}. 
The resonances shown are the 
$P_{33}(1232)$, $P_{11}(1440)$, $D_{15}(1520)$, $S_{11}(1535)$, $F_{15}(1680)$
and $F_{37}(1950)$. The nonresonant component is of electric dipole $E_1$
or $E_{0+}$ 
multipolarity as shown in the right panel and a mixture of  $E1$ and $M1$
multipolarity at higher energies}
\label{absorption-2}
\end{figure}

Figure  \ref{absorption-2} shows the total photoabsorption cross 
section of the proton.
This Figure has already been published in the 1970th \cite{armstrong72}
and there are
more precise data available. The advantage of this  Figure 
over these more recent data is that dispersion theory has been used 
to disentangle the total photoabsorption cross section into partial cross
sections which can be identified with the $s$-channel degrees of freedom of the
nucleon. In  Figure \ref{absorption-2} we see three main resonant states, 
the $P_{33}(1232)$
state which can be excited via a $M1(E2)$ transition and thus is 
responsible for the large paramagnetism of the nucleon, the $D_{13}(1520)$
state which can be reached via an $E1(M2)$ transition and the $F_{15}(1680)$
which 
can be reached via an $E2(M3)$ transition. These main resonant
states  correspond to the  three  oscillator shells with oscillator
quantum numbers $N=0$, $N=1$ and $N=2$ respectively. The further weaker
resonances are identified in the caption. 

In addition to the resonances, 
Figure \ref{absorption-2} contains a continuous nonresonant part. 
This continuous part of the spectrum 
corresponds to the nonresonant cross section where the low-energy part up to
about 500 MeV is mainly given by the $E_{0+}$ electric-dipole (E1)
single-pion photoproduction process. There are also minor magnetic-dipole (M1)
and mixed M1--E2 
single-pion photoproduction processes. At higher energies the photoabsorption
cross section is due to two-pion processes which mainly are of the type
$\gamma N\to \pi N^* \to N \pi\pi$ and, thus combine nonresonant and resonant
processes  during the absorption.

Though Figure \ref{absorption-2} is very useful as an overview 
\cite{schumacher07a} for
quantitative predictions of partial photoabsorption cross sections
information is more appropriate which has been obtained in recent time
from the study of photo-meson amplitudes \cite{drechsel07}. 
This will be explained in more
detail in  subsection 3.3.

\subsection{The level scheme of the nucleon}

\begin{figure}[h]
\begin{minipage}[b]{50mm}
\includegraphics[width=1.0\linewidth]{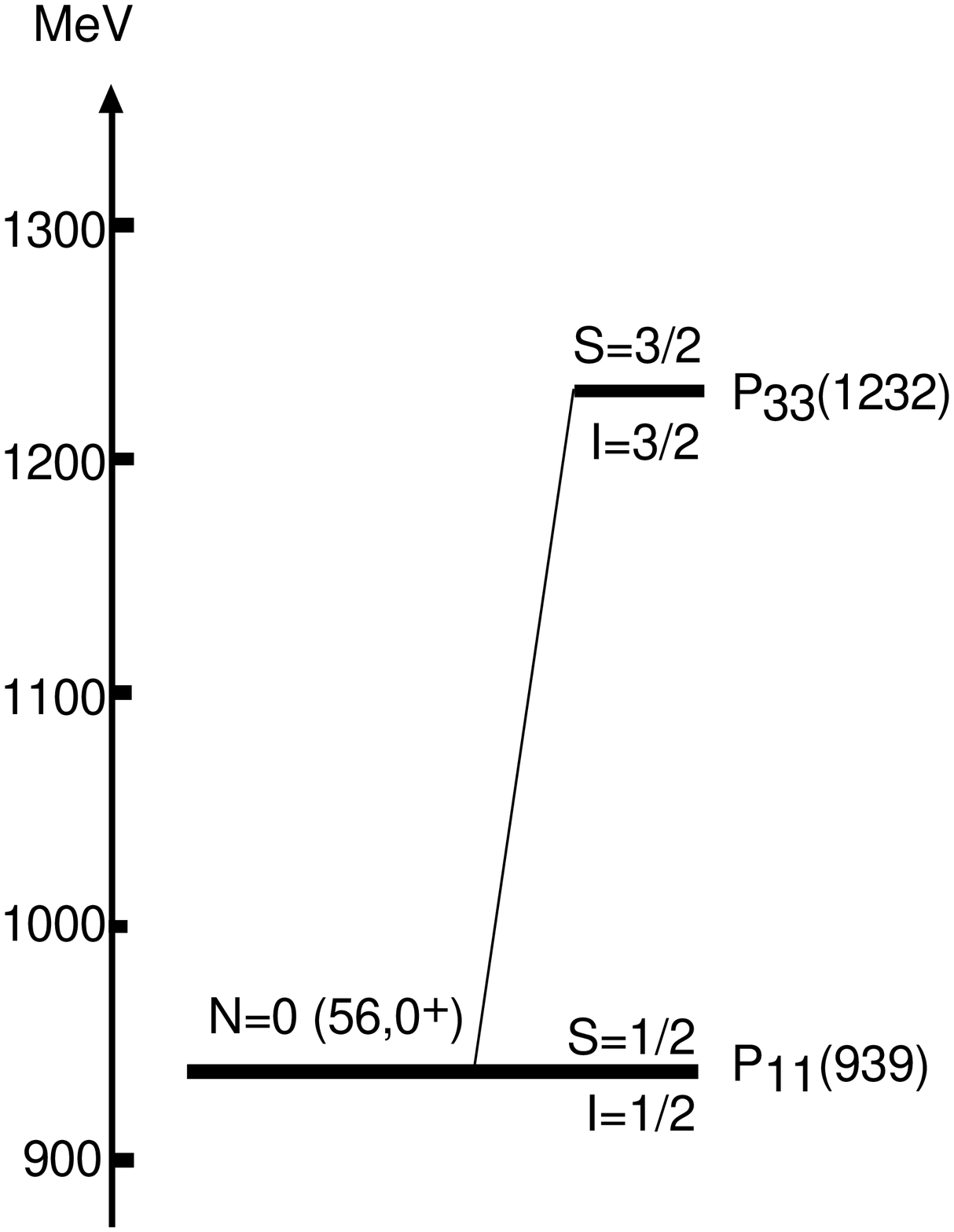}
\end{minipage}
\begin{minipage}[b]{50mm}
\includegraphics[width=1.0\linewidth]{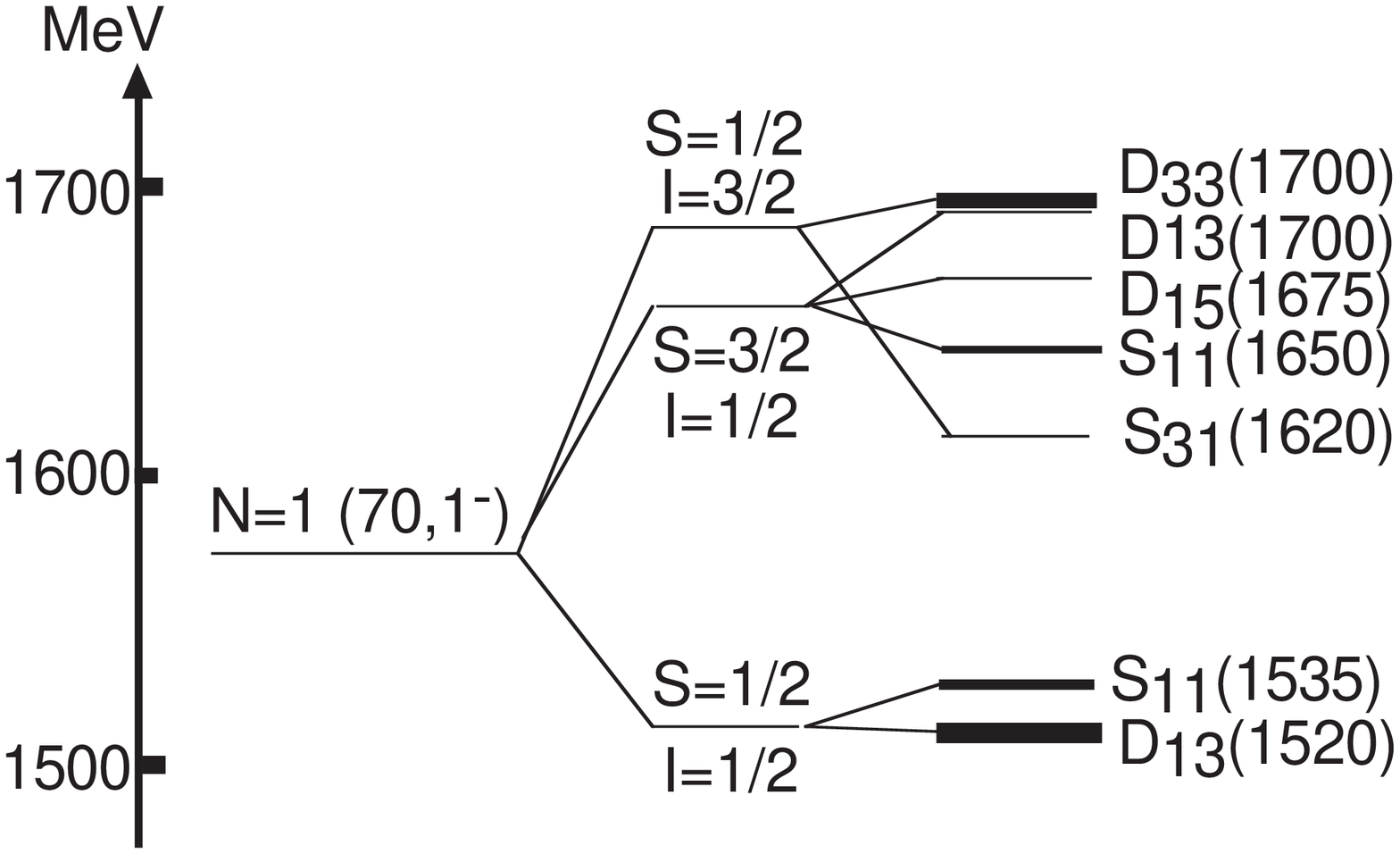}
\vspace{65mm}
\end{minipage}
\begin{minipage}[b]{50mm}
\includegraphics[width=1.00\linewidth]{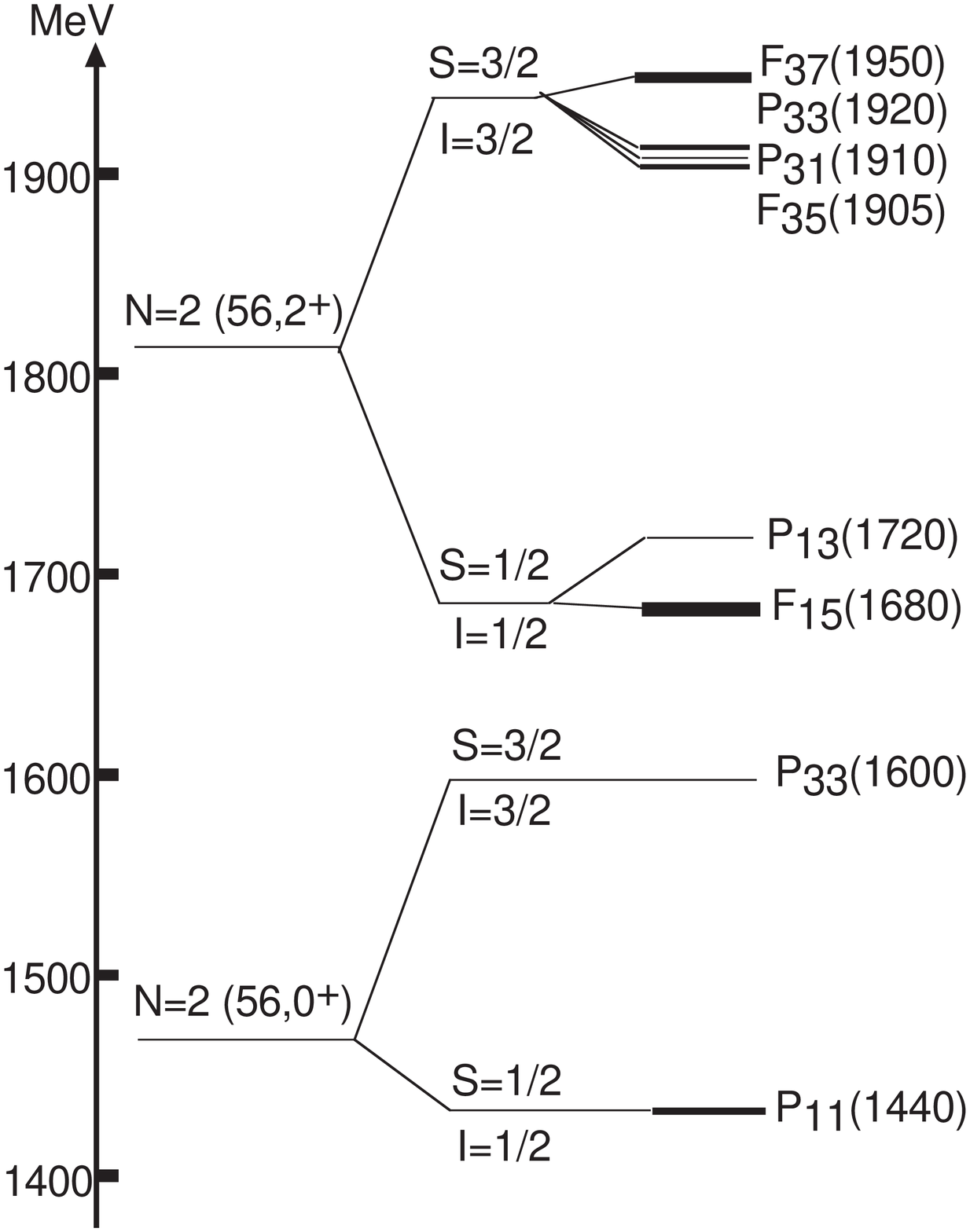}
\vspace{56mm}
\end{minipage}
\caption{The $N=0$ $(56,0^+_0)$ (left), $N=1$ 
$(70,1^-_1)$ (middle), and $N=2$  $(56,0^+_2)$, $(56,2^+_2)$ (right)
partial level schemes of the nucleon} 
\label{secondshell}
\end{figure}

Figure \ref{absorption-2} shows only the strongest resonances located below 2
GeV excitation energy. Therefore, it is of interest to present a complete
overview over the level scheme for this energy range.
Figure \ref{secondshell} shows the level scheme of the nucleon
corresponding to the first three HO shells. 
In total four resonance regions are formed where resonant states with
especially large strengths are found. Quantitative information on the strength
is given in \cite{schumacher09}  in terms of integrated photoabsorption cross
sections. 
In Figure \ref{secondshell}
the strong states are depicted  by thick
bars.  The first resonance region
is solely due to the $P_{33}(1232)$ state  which as a $I=3/2$ state
is of equal photon-excitation
strength for  the proton and the neutron. The second resonance
region is due to the $D_{13}(1520)$ state and to a lesser extent to the 
$S_{11}(1535)$ state. 
The third resonance region consists of the $F_{15}(1680)$
and the $D_{33}(1700)$ states. For  the proton the two contributions are of
approximately equal strength,  whereas for the neutron
according to the general rule for
$I=1/2$ states based on the quark model the $F_{15}(1680)$ contribution 
is largely suppressed. 
The fourth
resonance is due to the $F_{37}(1950)$ state which is of equal strength for the
proton and the neutron.

\subsection{Predicted  photoabsorption cross sections}

For the prediction of resonant cross sections in principle 
a direct calculation 
from the CGLN amplitudes is possible, provided the one-pion branching 
$\Gamma_\pi/\Gamma_r$ of the resonances is taken into account where necessary.
However, as shown in  \cite{schumacher09}, a much more
precise procedure is available. This method makes use of the parameters
of the resonances extracted from the CGLN data \cite{drechsel07}.

For the precise prediction
of $\sigma^{n}_T =\frac12\,
(\sigma^n_{3/2}+\sigma^n_{1/2})$ we use the Walker parameterization of resonant
states where the cross sections are represented in terms of Lorentzians
in the form
 \begin{equation}
I=I_r\left(\frac{k_r}{k}\right)^2\frac{W^2_r\,\Gamma(q)\Gamma^*_\gamma(k)}
{(W^2-W^2_r)^2 +W^2_r\,\Gamma^2(q)}
\label{peakcross1}
\end{equation}
where $W_r$ is the resonance energy of the resonance. The functions
$\Gamma(q)$ and $\Gamma^*_\gamma(k)$
are chosen such that a precise representation of the shapes
of the resonances are obtained. The appropriate parameterizations and the 
related references are given in \cite{schumacher09}.
Furthermore, some consideration given in \cite{schumacher09}
shows that the peak cross section can be expressed through the resonance
couplings in the following form
\begin{equation}
I_r=\frac{2\,m}{W_r \Gamma_r}\left[|A_{1/2}|^2+|A_{3/2}|^2\right].
\label{peakcross2}
\end{equation}
For the total
widths $\Gamma_r$ and the resonance couplings 
precise data are available.
Tabulations of values adopted for the present purpose are given in
\cite{schumacher09}.

\begin{figure}[ht]
\centering\includegraphics[width=0.47\linewidth]{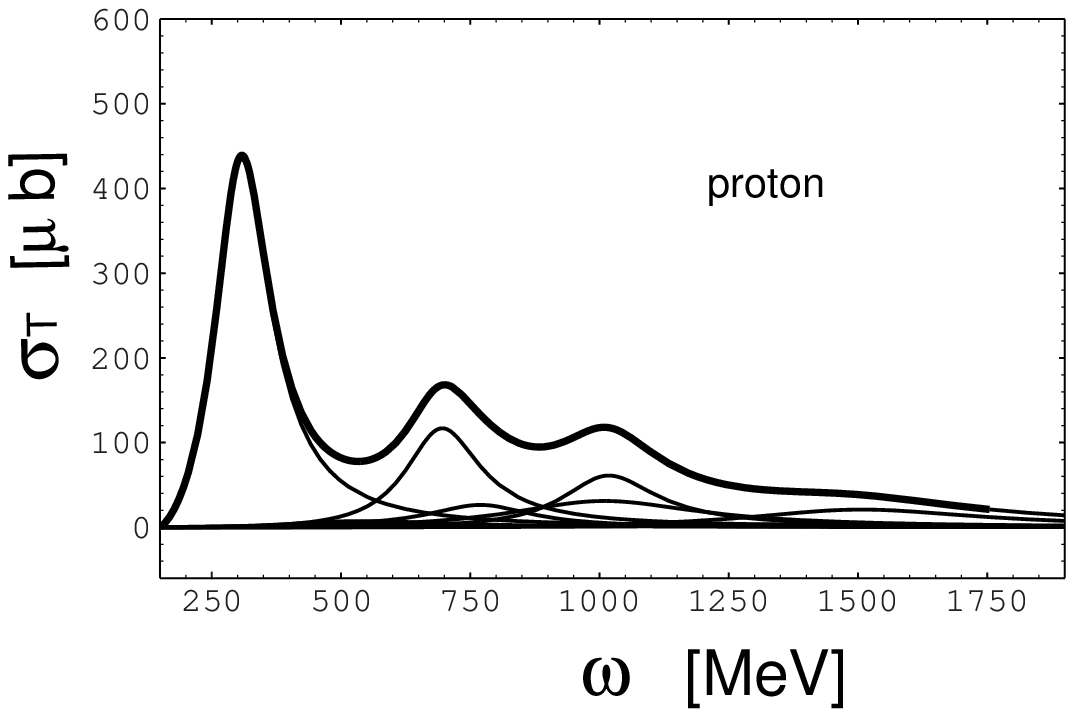}
\centering\includegraphics[width=0.47\linewidth]{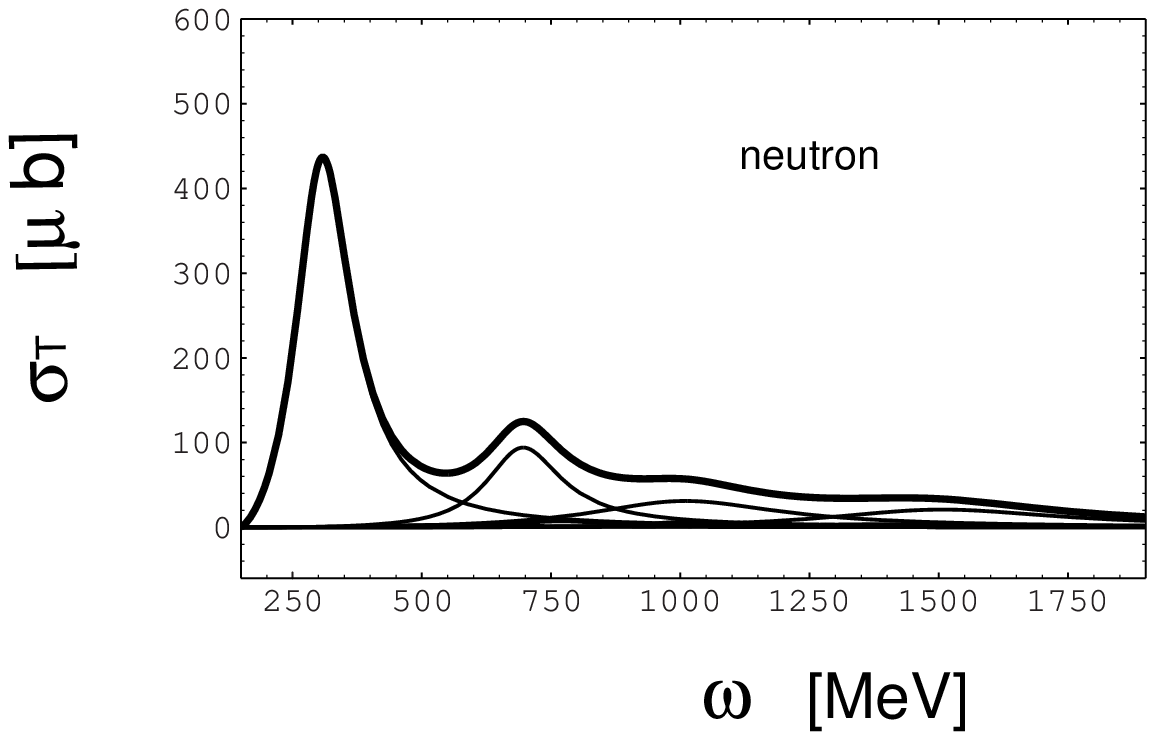}
\caption{Resonant part of the helicity independent
 photo-absorption cross section for the
  proton and the neutron. Thick line: Sum of all resonances. Thin lines:
Contributing single resonances. The energies and strengths corresponding to
  these resonances are given in Tables 7 and 8 of \cite{schumacher09}.}
\label{coordinates1}
\end{figure} 
\begin{figure}[ht]
\centering\includegraphics[width=0.47\linewidth]{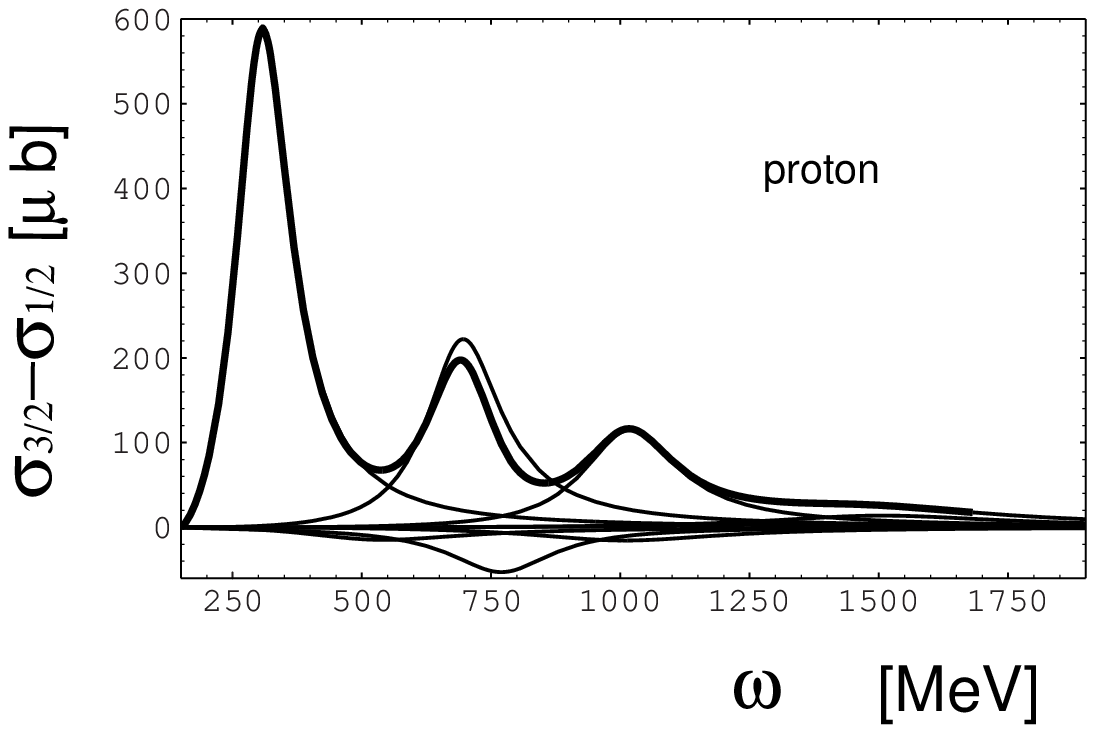}
\centering\includegraphics[width=0.47\linewidth]{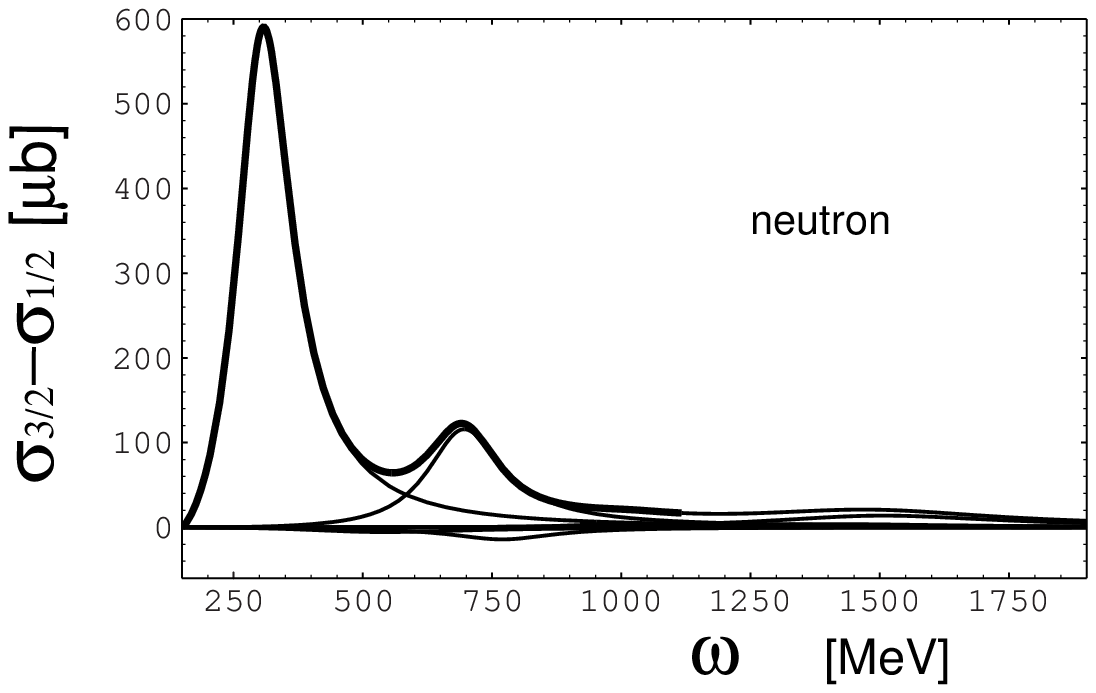}
\caption{Resonant part of the helicity dependent
 photo absorption cross section for the
  proton and the neutron. Thick line: Sum of all resonances. Thin lines:
Contributing single resonances. The energies and strengths corresponding to
  these resonances are given in Tables 7 and 8 of \cite{schumacher09}.
The signs of the cross sections are given in Table 3 of \cite{schumacher09}.
}
\label{coordinates2}
\end{figure}

The cross-section difference $\sigma_{3/2}-\sigma_{1/2}$ can also be
calculated directly  from the CGLN amplitudes. But also for this difference we 
propose a more appropriate procedure.
This procedure makes use of the fact
that precise values for total cross sections $\sigma_T$ are easier to obtain
than for the cross section differences $\sigma_{3/2}-\sigma_{1/2}$. Therefore,
we start from the relation
\begin{equation}
\sigma^{n}_{3/2}-\sigma^{n}_{1/2}=A_n\,\frac12\,
(\sigma^n_{3/2}+\sigma^n_{1/2}) =A_n\,\sigma^n_T
\label{crossansatz}
\end{equation}
where $n$ refers to the different resonant states.
The results obtained for $\sigma_T$ and $\sigma_{3/2}-\sigma_{1/2}$ 
from the present procedure are shown in Figures \ref{coordinates1}
and \ref{coordinates2}.
The resonance couplings $A_{3/2}$ and $A_{1/2}$, the widths $\Gamma_r$
and the scaling factors $A_n$ entering into (\ref{crossansatz}) are tabulated
in \cite{schumacher09}.

In Figure \ref{coordinates1} we show the total resonant
cross sections for the individual resonances of the proton and the neutron.
Four resonance regions are clearly visible in both cases, however with the 
difference that
for the neutron the third resonance region
is largely suppressed. This difference may
be traced back to the $F_{15}(1680)$ resonance which is smaller by a factor 
of 10 for the neutron as compared with the proton.
Figure  \ref{coordinates2} shows the corresponding cross section differences
$\sigma_{3/2}-\sigma_{1/2}$. Here again the line obtained for the 
$F_{15}(1680)$ resonance is largely suppressed in case of the neutron.

\subsection{Predicted  difference of helicity dependent cross sections 
 compared with experimental data \label{predicteddifference}}

Recent measurements at MAMI (Mainz) and ELSA (Bonn) have led to very valuable
data for the helicity dependence of the total photo-absorption cross section
of the proton and the neutron
\cite{GDH1,GDH2,GDH3,GDH4,GDH5,GDH6,GDH7,GDH8,GDH9,GDH10,GDH11,GDH12,GDH13,GDH14}. In the following we will compare experimental data obtained for
the proton with the predictions of the present approach. We restrict the
discussion to the proton because of the higher precision of the experimental
data as compared to the neutron.

\begin{figure}[h]
\includegraphics[width=0.45\linewidth]{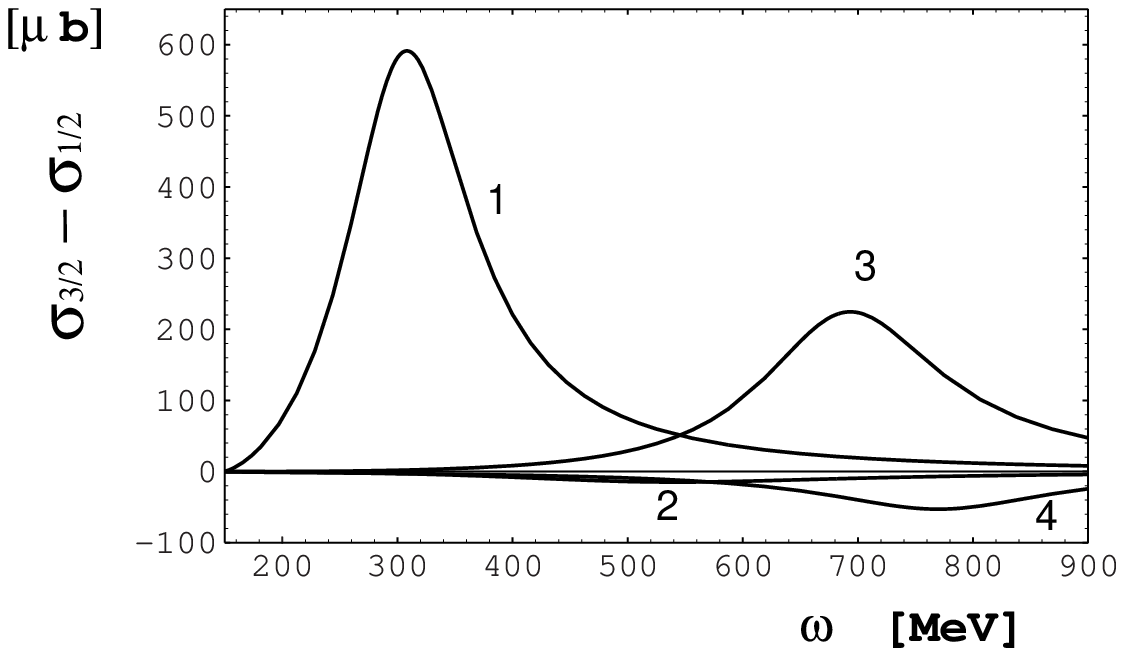}
\includegraphics[width=0.45\linewidth]{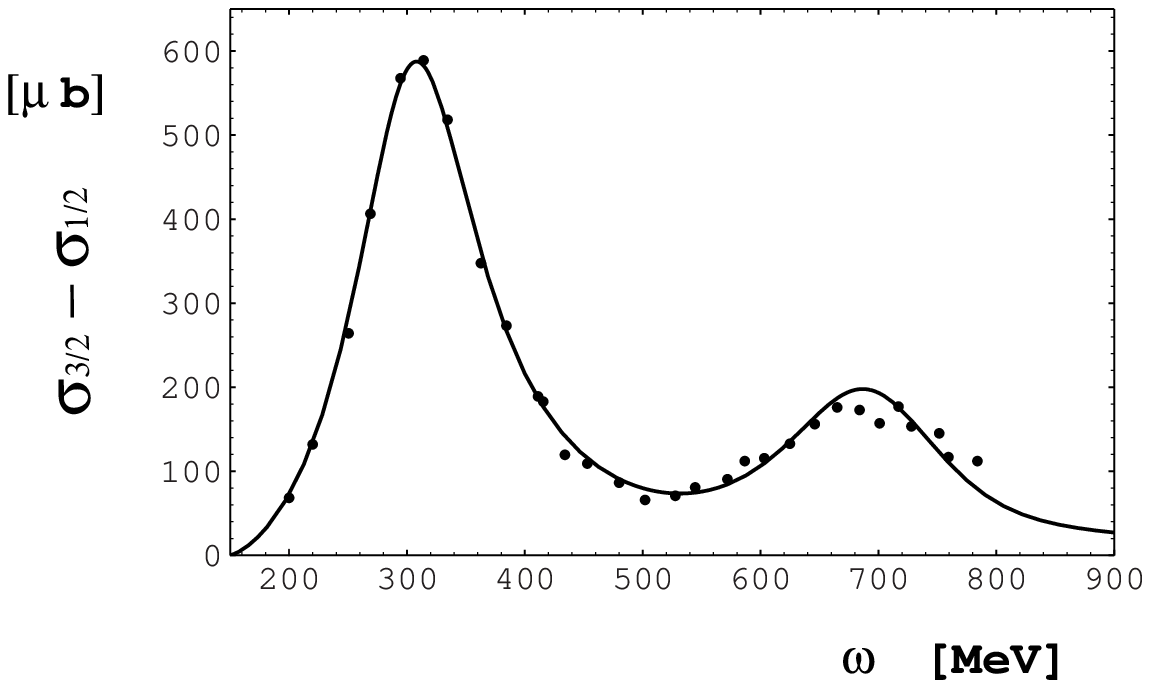}
\caption{Predicted  difference of helicity dependent 
cross sections for the proton
in the first and second resonance region compared with experimental data. 
The experimental data are taken from
\cite{GDH2}. The nonresonant contribution is eliminated from the figure by
a procedure described in the text. The contributions shown in the left panel
 are the resonances $P_{33}(1232)$ (1), $P_{11}(1440)$ (2),
$D_{13}(1520)$ (3), and $S_{11}(1535)$ (4).
}
\label{MAMI}
\end{figure}

In Figure  \ref{MAMI} we discuss experimental data obtained at MAMI (Mainz).
In order to clearly demonstrate the resonant structure of the 
first and the second resonance we have eliminated the  
nonresonant contribution from the figure and only keep the resonant
contribution. The baseline of the resonant contribution now is the abscissa.
Technically this has been achieved by adding the predicted nonresonant
contribution  to  the experimental data.     
The $P_{33}(1232)$ and
$D_{13}(1520)$ resonances make strong positive contributions whereas 
the $P_{11}(1440)$
and  $S_{13}(1535)$ resonances
make small negative contributions. During the fitting
procedure it was found out that slight shifts of some of the parameters within
their margins of errors led to an improvement of the fit to the experimental
data. For the $P_{33}(1232)$ resonance these shifts are
1232 MeV $\to$ 1226 MeV for the position 
\begin{figure}[h]
\includegraphics[width=0.45\linewidth]{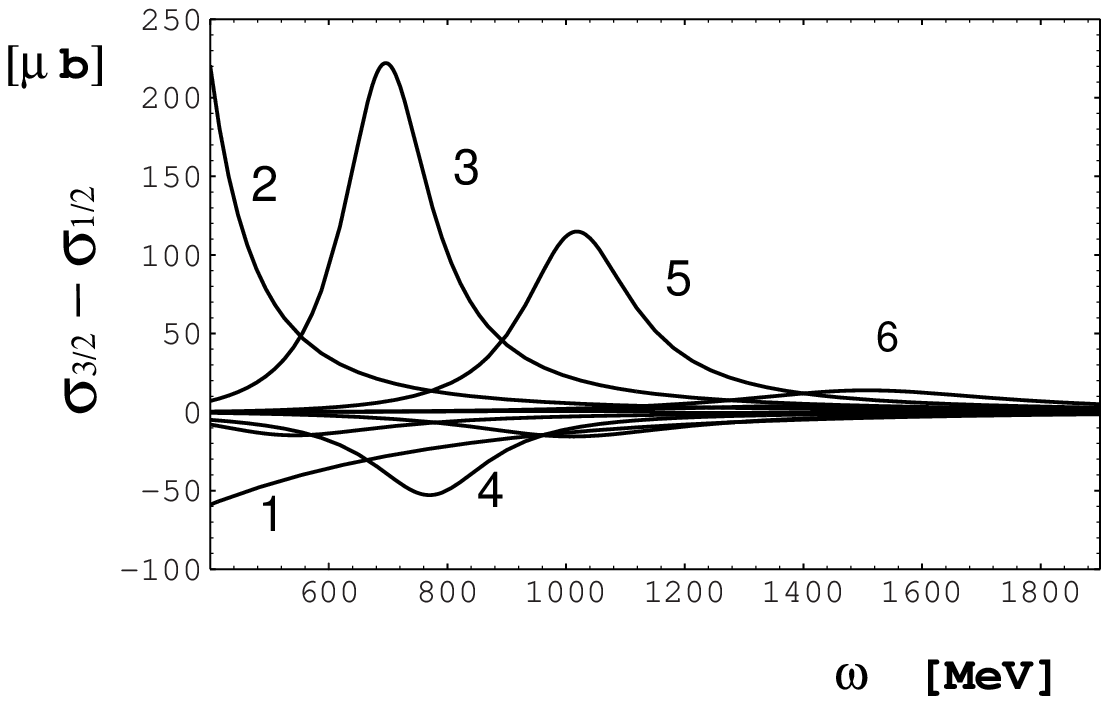}
\includegraphics[width=0.45\linewidth]{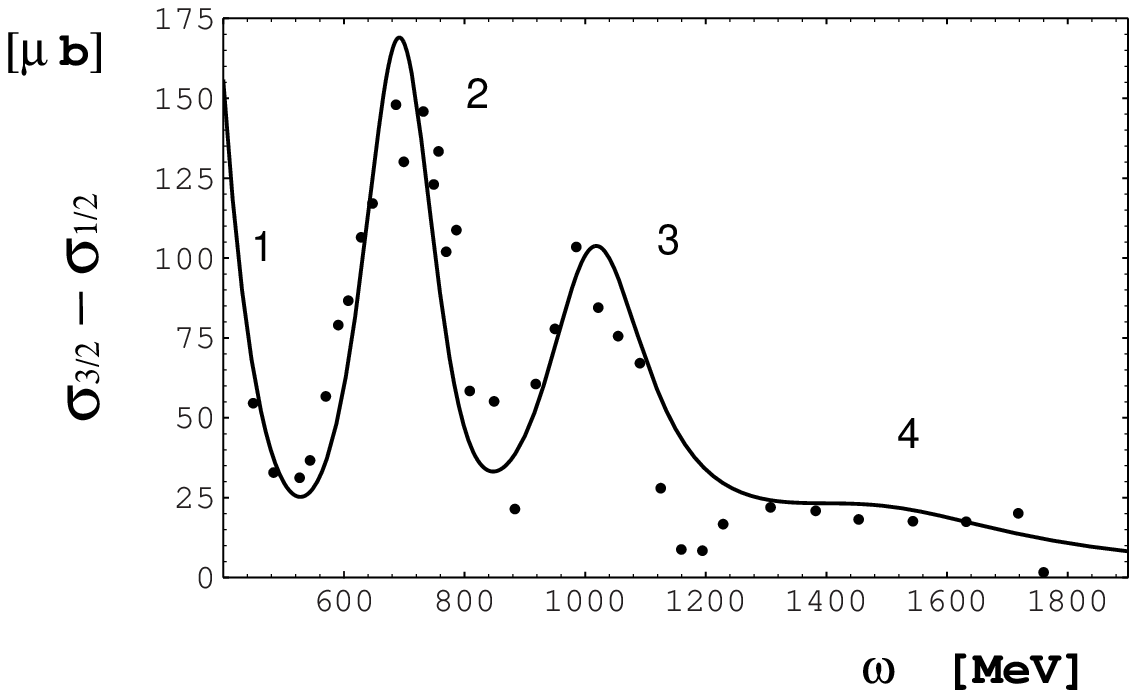}
\caption{Predicted  difference of helicity dependent 
cross sections for the proton
in the second, third and fourth  resonance region compared with experimental
data. The experimental data are taken from \cite{GDH7}. 
The contributions shown in the left panel
 are the tail of the  $P_{33}(1232)$ 
resonance (2), the nonresonant contribution (1),
the  $P_{11}(1440)$, $D_{13}(1520)$ (3),  $S_{11}(1535)$ (4), 
$S_{11}(1650)$,
$D_{15}(1675)$, $F_{15}(1680)$ (5), $D_{33}(1700)$, $F_{35}(1905)$ 
and $F_{37}(1950)$ (6)
resonances. The numbers given in the right panel denote the resonance regions
1--4.
}
\label{ELSA}
\end{figure}
of the peak, 130 MeV $\to$ 120 MeV for the width of the resonance and 
1.26 $\to$ 1.36 for the quantity $A_n$ defined in Eq. (\ref{crossansatz}).
This means that the integrated cross section and  also the
integrated  energy-weighted cross section remain constant.
For the $D_{13}(1520)$ resonance the shifts were 1520 $\to$ 1490 for 
the position
of the peak, 120 MeV $\to$ 130 for the width of the resonance. All other
parameters remained unmodified compared to the predictions given in 
\cite{schumacher09}.

Figure \ref{ELSA} differs from Figure \ref{MAMI} by the fact that the
nonresonant contribution is not eliminated from the Figure
but included into the theoretical curve. Indeed, the dip between
the first and the second resonance shown in the right panel
of Figure \ref{ELSA} is strongly influenced by the nonresonant cross sections
represented by curve 1 in the left panel. In the left panel only the stronger
resonances have been identified through a number, though all the relevant
resonances have been shown. As in Figure \ref{MAMI} some shifts of parameters
have been tested  in order to possibly improve on the fit to the experimental
data. The only shift of some relevance was the generation of a small negative 
contribution from the $D_{33}(1700)$ resonance by shifting the quantity 
$A_n$ from $0$ $\to$  $-0.5$.

The importance of the findings made in  Figures \ref{MAMI} and  \ref{ELSA} 
is that  our procedure of  applying  the Walker parameterization
to the helicity dependent cross section difference 
$(\sigma_{3/2}-\sigma_{1/2})$ has been tested and found valid. This makes it
possible to apply this procedure also to other problems as there is
the prediction and interpretation  of the $s$-channel contribution 
to the forward and backward angle spin-polarizabilities $\gamma_0$ and  
$\gamma_\pi$, respectively.
Furthermore, we have clearly demonstrated that the upper part of the cross
section in the right panel of Figure  \ref{ELSA} is due to the $F_{37}(1950)$
resonance as anticipated  before \cite{GDH7}.

\subsection{Nonresonant excitation of the nucleon} 

\begin{figure}[h]
\centering\includegraphics[width=0.45\linewidth]{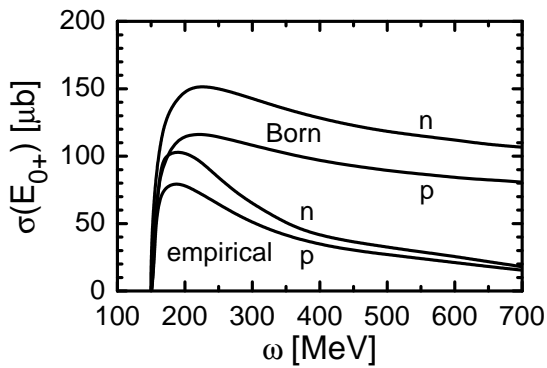}
\caption{Photoabsorption cross section $\sigma(E_{0+})$
due to $s$-wave single pion photoproduction for the proton (p) and the
neutron (n). The two upper  curves have been calculated in Born approximation.
The two lower  curves are empirical results  
\cite{hanstein98,drechsel99,drechsel07}
and extrapolations between 500 and 700 MeV}
\label{e0plus}
\end{figure}
In Figure \ref{e0plus}
the experimental $E_{0+}$ cross section is shown
 together with the prediction obtained in the Born approximation.
We see that in both cases the cross sections for the neutron are larger
than the cross sections for the proton. For the Born approximation this has
been discussed in the textbook of Ericson and Weise (see p. 286 of 
\cite{ericson88}). The explanation is that for meson photoproduction
the electric dipole amplitude $E_{0+}$ is larger for the reaction 
$\gamma n \to \pi^- p$ than for the reaction $\gamma p \to \pi^+ n$
because of the larger electric dipole moment in the final state. Indeed the 
ratio of the two cross sections given for the Born approximation
can be obtained  from ratio of the $\pi^- p$ and $\pi^+ n$ dipole moments. 
The  difference in dipole moments which also shows up in the empirical $E_{0+}$
cross sections leads to an explanation for the fact that the neutron has a
larger electric polarizability than the proton.
The empirical cross section shown in Figure \ref{e0plus} above 
$\omega=500$ MeV  are 
extrapolations of data given in \cite{hanstein98,drechsel99,drechsel07}.
The extrapolation was necessary because this cross section  is purely 
nonresonant only up to $\omega= 500$ MeV.  Above this energy the
interference of cross sections from nucleon resonances is observed. The
guidance for the extrapolation is taken from the Born approximation showing
that the cross section is almost a straight line at energies above 450 MeV.

\subsection{Diagrammatic representation of Compton scattering}

\begin{figure}[h]
\begin{center}
\includegraphics[width=0.5\linewidth]{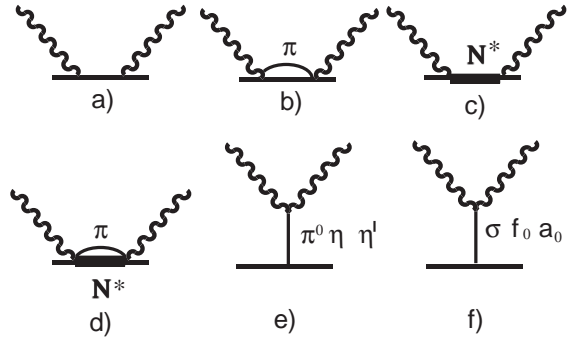}
\end{center}
\caption{Graphs  of nucleon Compton scattering. a) Born term,
b) single-pion photoproduction, c) resonant excitation of the nucleon
and d) two-pion photoproduction. 
The crossed graphs 
accompanying the graphs a) to d) are not shown. The $t$-channel graphs in e)
and f) correspond to pointlike singularities due to the pseudoscalar meson
$\pi^0$, $\eta$ and $\eta'$ and the scalar mesons $\sigma(600)$, $f_0(980)$
and $a_0(980)$. These  graphs are shown for illustration whereas
quantitative predictions are based on dispersion theory.}
\label{comp-graph}
\end{figure}
The method of predicting  polarizabilities is based on dispersion theory
applied to partial photoabsorption cross-sections. 
Nevertheless, it is of interest
for illustration to show the Compton scattering process in terms of
diagrams as carried out in Figure \ref{comp-graph}.
We see that in addition to the nonresonant single-pion channel (b)
and the resonant channel (c) we have three further contributions.
Graph (d) corresponds to Compton scattering via the transition
$\gamma N\to \Delta \pi$ to the intermediate state.   
This represents  the main component of those photomeson processes 
where two pions are in the final state. Graph e) is the well-known
pseudoscalar $t$-channel graph where in addition to the dominant
$\pi^0$ pole contribution also the small contributions due to $\eta$
and $\eta'$ mesons are taken into account. Graph f) depicts the
scalar counterpart of the pseudoscalar pole term e). At  first sight
these scalar pole-terms appear inappropriate because the scalar particles
and especially the $\sigma$ meson have a large width when observed on shell.  
However, this consideration overlooks that in a Compton scattering process
the scalar particles are intermediate states of the scattering process
where the broadening due to the $\pi\pi$ final state is not effective. 
This leads to the conclusion that the scalar pole terms indeed provide the
correct description of the scattering process.
Further information is given in subsection 5.4.

\subsection{Predicted polarizabilities obtained for the $s$-channel}

Using the sum rules given in Eqs. (\ref{baldin}) to 
(\ref{polar7})  and applying them to
the photoabsorption cross sections given in subsections 3.3 to 3.5 the
polarizabilities  and spin-polarizabilities given in Table 
\ref{polresults1} are obtained.  Lines 2-12 contain the contributions of the
nucleon resonances, with the sum of the resonant contributions given in line
13. The nonresonant electric and magnetic single-pion contributions are given
in lines 14-16, and the contribution due to the two-pion channel in line
17. Line 18 contains the total $s$-channel contribution.  

 \begin{table}[h]
\caption{Resonant (lines 2--12),  single-pion nonresonant (lines 14--16) and
two-pion (line 17) components of the polarizabilities. Line 13 contains the
total resonant components  and line 18 the total $s$-channel component.
The electric and magnetic
  polarizabilities are in units of $10^{-4}$fm$^3$, the spin
  polarizabilities  in units of $10^{-4}$fm$^4$.}
\begin{center}
\begin{tabular}{l|ll|ll|ll|ll}
\hline
&$\alpha_p$&$\beta_p$&$\alpha_n$&$\beta_n$&
$\gamma^{(p)}_0$&$\gamma^{(n)}_0$&$\gamma^{(p)}_\pi$&$\gamma^{(n)}_\pi$\\ 
\hline
$P_{33}(1232)$&$-1.07$&$+8.32$&$-1.07$&$+8.32$ &
$-3.03$&$-3.03$&$+5.11$&$+5.11$\\         
$P_{11}(1440)$&$-0.02$&$+0.14$&$-0.01$&$+0.05$&
$+0.05$&$+0.02$&$-0.10$&$-0.04$\\
$D_{13}(1520)$&$+0.68$&$-0.16$&$+0.55$&$-0.13$&
$-0.14$&$-0.07$&$-0.39$&$-0.20$\\
$S_{11}(1535)$&$+0.21$&$-0.05$&$+0.06$&$-0.01$&
$+0.05$&$+0.01$&$+0.13$&$+0.04$\\
$S_{11}(1650)$&$+0.05$&$-0.01$&$+0.00$&$-0.00$&
$+0.01$&$+0.00$&$+0.03$&$+0.00$\\
$D_{15}(1675)$&$+0.01$&$-0.00$&$+0.08$&$-0.02$&
$-0.00$&$-0.00$&$-0.00$&$-0.01$\\
$F_{15}(1680)$&$-0.07$&$+0.25$&$-0.01$&$+0.03$&
$-0.02$&$-0.00$&$+0.13$&$+0.01$\\
$D_{33}(1700)$&$+0.25$&$-0.07$&$+0.25$&$-0.07$&&&$-0.00$&$-0.00$\\
$F_{35}(1905)$&$-0.01$&$+0.02$&$-0.01$&$+0.02$&&&$+0.01$&$+0.01$\\
$P_{33}(1920)$&$-0.01$&$+0.02$&$-0.01$&$+0.02$&&&$+0.01$&$+0.01$\\
$F_{37}(1950)$&$-0.03$&$+0.10$&$-0.03$&$+0.10$&&&$+0.01$&$+0.01$\\
\hline
sum-res.&$-0.01$&$+8.56$&$-0.20$&$+8.31$&
$-3.08$&$-3.07$&$+4.94$&$+4.94$\\
\hline
$E_{0+}$&$+3.19$&$-0.34$&$+4.07$&$-0.43$&
$+2.47$&$+3.18$&$+3.75$&$+4.81$\\
$M^{(3/2)}_{1-}$&$-0.04$&$+0.35$&$-0.04$&$+0.35$&
$-0.11$&$-0.11$&$-0.18$&$-0.18$\\
$(M,E)^{(1/2)}_{1+}$&$-0.06$&$+0.47$&$-0.21$&$+1.44$&
$+0.14$&$+0.38$&$+0.24$&$+0.66$\\
$\gamma N\to \pi\Delta$&$+1.4$&$+0.4$&$+1.5$&$+0.4$&
$\,\,\,\,\,0.00$&$\,\,\,\,\,0.00$&$-0.28$&$-0.23$\\
\hline
$s$-channel&$+4.48 $&$+9.44 $&$+5.12 $&$+10.07 $&
$-0.58 $&$+0.38 $&$+8.47  $&$+10.00 $\\
\hline
\hline
\end{tabular}
\label{polresults1}
\end{center}
\end{table}
The forward-angle spin-polarizabilities have the peculiarity that the proton
and neutron values have different signs and both are very close to zero.
Therefore these values should be given with an error. 
Taking into account the well known rules of error analysis one obtains
\begin{equation}
\gamma^{(p)}_0=-0.58\pm 0.20,\quad \gamma^{(n)}_0=+0.38\pm 0.22.
\label{res}
\end{equation} 

For the proton the present result may be compared with the most recent
previous evaluation \cite{pasquini10}. The main result of this evaluation
is $\gamma^{(p)}_0=-0.90\pm 0.08 \pm 0.11$. Other results based on different 
photomeson analyses are $\gamma^{(p)}_0= -0.67$ (HDT), $-0.65$ (MAID), 
$-0.86$ (SAID) and $-0.76$ (DMT). In view of the fact that different 
data sets  have been used in these analyses the consistency of these results
and the agreement with our result
appears remarkably good. 
 A comparison is also possible
with the analysis of Drechsel et al. \cite{drechsel98} based on CGLN
amplitudes and dispersion theory. The numbers obtained are 
$\gamma^{(p)}_0=-0.6$ and $\gamma^{(n)}_0=+0.0$ based on the HDT
parameterization. The result obtained in
\cite{drechsel98} for the proton is in close agreement with our result.
The result obtained in \cite{drechsel98}
for the neutron confirms our prediction  that the quantity $\gamma^{(n)}_0$
has the tendency of being shifted towards positive values.
In \cite{drechsel98}
also a detailed comparison with chiral perturbation theory is given which
should not be  repeated here.

\section{S-matrix and invariant amplitudes}

For the following part of the present paper it is necessary to outline
the theoretical tools on a wider basis. These tools are the S-matrix
for nucleon Compton scattering and the invariant amplitudes.

\subsection{S-matrix}
The amplitude $T_{fi}$ for Compton scattering
\begin{equation}
\gamma(k) N(p) \to \gamma'(k') N(p')
\label{D1}
\end{equation}
is related to the $S$-matrix of the reaction through the relation
\begin{equation}
\langle f|S-1|i \rangle= i (2\pi)^4 \delta^4(k+p-k'-p')T_{fi}.
\label{D2}
\end{equation}
The quantities $k=(\omega,{\bf k})$, $k'=(\omega',{\bf k}')$,
$p=(E,{\bf p})$, $p'=(E',{\bf p'})$ are the four momenta of the photon
and the nucleon in the initial and final states, respectively,
related to the Mandelstam variables via 
\begin{equation}
s=(k+p)^2, \quad t=(k-k')^2, \quad u=(k-p')^2.
\end{equation}

The scattering amplitude $T_{fi}$ may be expressed on an
orthogonal basis suggested by Prange \cite{prange58} by 
means of six invariant amplitudes $T_k(\nu,t)$ leading to the general
form \cite{lvov97} 
\begin{eqnarray}
T_{fi}&=\bar{u'} e'{}^{*\mu} \Big[
-\frac{P'_\mu P'_\nu}{P'{}^2}(T_1+(\gamma \cdot K)T_2) - 
\frac{N_\mu N_\nu}{N^2}
(T_3+(\gamma \cdot K)T_4)\nonumber\\ &+i \frac{P'_\mu N_\nu -P'_\nu
  N_\mu}{P'{}^2K^2}\gamma_5
T_5 + i\frac{P'_\mu N_\nu +P'_\nu N_\mu}{P'^2K^2}
\gamma_5 (\gamma \cdot K) T_6\Big] e^\nu u. \label{Tif}
\end{eqnarray}
In (\ref{Tif}) $u'$ and $u$ are the Dirac spinors of the final and the
initial nucleon, $e'$  and $e$  are the
polarization 4-vectors of the final and the initial photon, and 
$\gamma_5=-i\gamma_0\gamma_1\gamma_2\gamma_3$.  The 4-vectors
$P'$, $K$ and $N$ together with the vector $Q$ are orthogonal and
are expressed in terms 
of the 4-momenta $p'$, $k'$
and $p$, $k$ of the final  and initial nucleon and photon,
respectively,
by
\begin{eqnarray}
&&K=\frac12 (k+k'),\quad P'=P-\frac{K(P\cdot K)}{K^2},\quad 
N_\mu=\epsilon_{\mu\nu\lambda\sigma}  P'{}^\nu Q^\lambda
K^\sigma,\nonumber\\
&&P=\frac12 (p+p'),\quad Q=\frac12 (k'-k)=\frac12 (p-p')\label{KPN}
\end{eqnarray}
where $\epsilon_{\mu\nu\lambda\sigma}$ is the antisymmetric tensor with
$\epsilon_{0123}=1$. The amplitudes  $T_k$ are functions of the two
variables $\nu$ and $t$  with
\begin{equation}
\nu=\frac{s-m^2+t/2}{2m}
\label{nu}
\end{equation}
 and $m$ is the mass of the nucleon.
The normalization of the amplitude $T_{fi}$ is determined by
\begin{equation}
\bar{u}u=2m, \quad \frac{d\sigma}{d\Omega}=\frac{1}{64 \pi^2
  s}\sum_{\rm spins}|T_{fi}|^2.
\label{NORM}
\end{equation}
It follows from crossing symmetry of $T_{fi}$, that $T_{1,3,5,6}$ and
$T_{2,4}$
are even and odd functions of $\nu$, respectively.

The amplitudes $T_k(\nu,t)$  do not have
kinematic singularities, but there are kinematic constraints, which
arise from the vanishing of $P'{}^2$, $N^2$ and $P'{}^2K^2$ in the
dominators of the decomposition (\ref{Tif}) at certain values of $\nu$
and $t$. The kinematic constraints can be removed by introducing
linear combinations of the amplitudes $T_k(\nu,t)$. This problem has first
been satisfactorily solved 
 by Bardeen and Tung \cite{bardeen68}. However, the Bardeen and Tung 
amplitudes contain the inconvenience that part of them are even functions of
$\nu$ and part of them odd functions. This inconvenience has been removed by
L'vov \cite{lvov81} so that these latter amplitudes now have become standard.

\subsection{Invariant amplitudes}

The linear combinations  introduced by 
L'vov \cite{lvov81} are 
\begin{eqnarray}
&&A_1=\frac{1}{t}[T_1+T_3+\nu (T_2+T_4) ], \nonumber \\
&&A_2=\frac{1}{t}[2T_5 +\nu (T_2+T_4)], \nonumber\\
&&A_3=\frac{m^2}{m^4-su}\left[T_1-T_3 - \frac{t}{4\nu}(T_2-T_4)\right],
\nonumber\\
&&A_4=\frac{m^2}{m^4-su}\left[2m
  T_6-\frac{t}{4\nu}(T_2-T_4)\right],\nonumber\\
&&A_5=\frac{1}{4\nu}[T_2+T_4],\nonumber\\
&&A_6=\frac{1}{4\nu }[T_2-T_4]. \label{T1-6}
\end{eqnarray}
The amplitudes $A_i(\nu,t)$ are even functions of $\nu$ and have no
kinematic singularities or kinematic constraints. They
have poles at zero energy because of contributions of the nucleon in
the intermediate state. These poles are contained in two Born diagrams
with the pole propagator $(\gamma\cdot p -m)^{-1}$ and on-shell
vertices $ \Gamma_\mu(p+k,p)=\gamma_\mu+[\gamma\cdot
  k,\gamma_\mu]\kappa/4m$,
where $\kappa=1.793 q- 1.913(1-q)$ is the nucleon anomalous magnetic
moment. Here the electric charge of the nucleon,
$q=\frac12 (1+\tau_3)=1$ or $0$ is introduced. The Born contributions to the
amplitudes $A_i$ have a pure pole form
\begin{equation}
A^{\rm B}_i(\nu,t)=\frac{m e^2 r_i(t)}{(s-m^2)(u-m^2)},
\label{Born}
\end{equation}
where e is the elementary electric charge ($e^2/4\pi = 1/137.04$)
and 
\begin{eqnarray}
&&r_1=-2q^2+(\kappa^2+2q\kappa)\frac{t}{4m^2},\quad
r_2=2q\kappa+2q^2+(\kappa^2+2q\kappa)\frac{t}{4m^2},\nonumber\\
&&r_3=r_5=\kappa^2+2q\kappa,\quad r_4=\kappa^2,\quad
r_6=-\kappa^2-2q\kappa-2q^2.
\label{residue}
\end{eqnarray}

\subsection{Lorentz invariant definition of polarizabilities}

The relations between 
the amplitudes $f$ and $g$ introduced in  Section 2.1
and the invariant amplitudes $A_i$ \cite{babusci98,lvov97}
 are
\begin{eqnarray}
&&f_0(\omega)= -\frac{\omega^2}{2\pi}\left[A_3(\nu,t)+ A_6(\nu,t)
\right],\quad\quad\quad
g_0(\omega)=\frac{\omega^3}{2\pi m}A_4(\nu,t), \label{T3}\\
&&f_\pi(\omega)=-\frac{\omega\omega'}{2\pi}\left(1+\frac{\omega\omega'}
{m^2}\right)^{1/2}\left[ 
A_1(\nu,t) - \frac{t}{4 m^2}A_5(\nu,t)\right],\label{T4}\\
&&g_\pi(\omega)=-\frac{\omega\omega'}{2\pi m}\sqrt{\omega\omega'}
\left[
A_2(\nu,t)+ \left(1-\frac{t}{4 m^2}\right)A_5(\nu,t)\right]
,\label{T5}\\
&&\omega'(\theta=\pi)=\frac{\omega}{1+2\frac{\omega}{m}},\,\,
\nu=\frac12 (\omega+\omega'),\,\, t(\theta=0)=0,
\,\,t(\theta=\pi)=-4\omega\omega.'
\label{T6}
\end{eqnarray}
For the electric, $\alpha$, and magnetic, $\beta$,  polarizabilities 
and the spin polarizabilities $\gamma_0$ and $\gamma_\pi$ for the
forward and backward directions, respectively, 
we obtain the relations
\begin{eqnarray}
&&\alpha+\beta = -\frac{1}{2\pi}\left[A^{\rm nB}_3(0,0)+ 
A^{\rm nB}_6(0,0)\right], \quad 
\alpha-\beta = -\frac{1}{2\pi}
\left[A^{\rm nB}_1(0,0)\right], \nonumber\\ 
&&\gamma_0= \frac{1}{2\pi m}\left[A^{\rm nB}_4(0,0)
\right], \quad\quad\quad\quad \quad\quad\quad\,\,\,
\gamma_\pi = -\frac{1}{2\pi m}
\left[A^{\rm nB}_2(0,0)+A^{\rm nB}_5(0,0) \right],
\label{T7}
\end{eqnarray}
showing that the invariant amplitudes may be understood as a 
generalization of the  polarizabilities.

According to Eqs. (\ref{T3}) to  (\ref{T7}) the following linear
combinations of invariant amplitudes are of special importance because 
they contain the physics of the four fundamental sum rules, { \it viz.}
the BEFT (\ref{T8}), LN (\ref{T9}), BL (\ref{T10})  
and GDH (\ref{T11}) sum rules, respectively:
\begin{eqnarray}
&& {\tilde A}_1(\nu,t)\equiv A_1(\nu,t)-\frac{t}{4m^2}A_5(\nu,t),\label{T8}\\
&& {\tilde A}_2(\nu,t)\equiv A_2(\nu,t)+\left(1-\frac{t}{4m^2}\right)
A_5(\nu,t),\label{T9}\\
&& {\tilde A}_3(\nu,t)  \equiv A_{3+6}(\nu,t)\equiv A_3(\nu,t)+ 
A_6(\nu,t), \label{T10}\\
&& {\tilde A}_4(\nu,t)   \equiv A_4(\nu,t).
\label{T11}
\end{eqnarray}

\section{From the Bernabeu-Ericson-FerroFontan-Tarrach 
(BEFT) sum rule to the $\sigma$-meson pole}

The introduction of the $\sigma$-meson pole dates back to the seminal
paper of L'vov et al. \cite{lvov97},  where for the first time a precise 
prediction of Compton differential  cross sections in the second
resonance region of the proton  for a large angular interval was achieved
The observation made in these investigations was that at  high energies
and large scattering angles the prediction of Compton differential cross
sections requires a large contribution from the scalar $t$-channel. In 
\cite{lvov97} it was  empirically shown that this contribution
can be represented by a $\sigma$-meson pole with a real mass $m_\sigma$
of about 600 MeV  in complete analogy to the
$\pi^0$-meson pole. However, there remained an important  difference 
between the two poles, because
the $\pi^0$-meson pole had a firm theoretical justification, whereas the 
$\sigma$-meson pole was considered as a  tool of computation without
a good theoretical basis. The reason was that in reactions like
$\gamma\gamma\to \sigma \to \pi\pi$ the $\sigma$-meson is represented by a
pole with a complex mass 
$\sqrt{s_\sigma}=M_\sigma-i\,\frac12 
\Gamma_\sigma$
located on the second Riemann sheet. This fact seems to contradict the
assumption of a pole located on the real $s$-axis.

An essential step forward was made in \cite{schumacher06} where it was shown
that the use of arguments contained in the work of Scadron et
al. \cite{delbourgo95} leads to a 
firm justification  of the $\sigma$-pole ansatz and to a quantitative
prediction of $(\alpha-\beta)^t$ and $m_\sigma$. The corresponding chain of
arguments are outlined in the following.

\subsection{The BEFT sum rule}

The BEFT sum rule 
\cite{bernabeu74,bernabeu77,guiasu76,guiasu78,budnev79,holstein94}
may be derived from the non-Born part of the
invariant amplitude (see \cite{lvov97,schumacher05})
\begin{equation}
{\tilde A}_1(s,u,t) \equiv A_1(s,u,t) -\frac{t}{4m^2}A_5(s,u,t)
\label{A-1-tilde}
\end{equation}
by applying the fixed-$\theta$ dispersion relation for
$\theta=180^\circ$.
Then we arrive at
\begin{eqnarray}
{\rm Re}{\tilde A}^{\rm nB}_1(s,u,t) &=& \frac{1}{\pi}{\cal P}
\int^\infty_{s_0}\left(\frac{1}{s'-s}+\frac{1}{s'-u}-\frac{1}{s'}\right)
{\rm Im}_s{\tilde A}_1(s',u',t')ds'\nonumber\\
&+& \frac{1}{\pi}{\cal P}\int^\infty_{t_0}{\rm Im}_t {\tilde A}_1(s',u',t') 
\frac{dt'}{t'-t}\,\, ,
\label{disprelA1}
\end{eqnarray}
where $s_0=(m+m_\pi)^2$, $t_0= 4 m^2_\pi$, $s'+u'+t'=2m^2$ and $s'u'=m^4$.
Using 
\begin{equation}
\alpha-\beta=-\frac{1}{2\pi}{\tilde A}^{\rm nB}_1(m^2,m^2,0)=(\alpha-\beta)^s+
(\alpha-\beta)^t
\label{alpha-beta-BEFT}
\end{equation}
we arrive at
\begin{eqnarray}
&&(\alpha - \beta)^s = - \frac{1}{2\pi^2} \int^\infty_{s_0}
\frac{s'+m^2}{s'-m^2}{\rm Im}_s {\tilde
  A}_1(s',u',t')\frac{ds'}{s'}\,\, ,\label{SR1}\\
&&(\alpha-\beta)^t=
-\frac{1}{2\pi^2}\int^\infty_{t_0}{\rm Im}_t
{\tilde A}_1(s',u',t') \frac{dt'}{t'}.
\label{SR}
\end{eqnarray}

\subsubsection{The $s$-channel part of the BEFT sum rule}

Using Eq. (\ref{SR1}) and    
the relation for ${\rm Im}f_\pi$ derived in Eq. (\ref{optical-5})
we arrive at the $s$-channel part of the BEFT sum rule given in 
Eq. (\ref{BEFT}).
The main contribution to $(\alpha-\beta)^s$ comes from the
photoproduction of $\pi N$ states. For these states, the cross-section
difference can be expressed in terms of the standard CGLN amplitudes
via 
\begin{eqnarray}
&&\left[ \sigma(\omega,\Delta P= {\rm yes}) -\sigma(\omega,\Delta P = {\rm no})
  \right]^{\pi N}= 4\pi\frac{q}{k}\sum^\infty_{k=0} (-1)^k
(k+1)^2\nonumber\\   && \hspace{2.cm} \times
\Big\{ (k+2)\left(|E_{k+}|^2- |M_{(k+1)-}|^2\right)
+ k\left(|M_{k+}|^2-|E_{(k+1)-}|^2\right)\Big\}^{\pi N},
\label{CGLN}
\end{eqnarray}
where $q=|{\bf q}|$ and $k=|{\bf k}|$  are the pion and photon 
momenta in the cm system, respectively. 

In (\ref{CGLN}) the quantity $\left[ \sigma(\omega,\Delta P= {\rm yes})
  -\sigma(\omega,\Delta P = {\rm no})\right] $ is expressed in terms of CGLN
amplitudes which where barely known when the first  studies of the
BEFT sum rule where carried out and remained of limited precision later on. 
A great improvement was possible in the
present investigation where the difference 
$\left[ \sigma(\omega,\Delta P= {\rm yes})
  -\sigma(\omega,\Delta P = {\rm no})\right]$ is calculated from partial cross
sections
as described in section 3.

\subsubsection{The $t$-channel part of the BEFT sum rule}

The imaginary part of the amplitude ${\tilde A}_1$ (Eq. \ref{A-1-tilde})
in the $t$-channel
can be found using the general unitarity relation
\begin{equation}
{\rm Im}_t T(\gamma\gamma\to N{\bar N})=\frac12 \sum_n (2\pi)^4
\delta^4(P_n-P_i)T(\gamma\gamma\to n)T^*(N{\bar N}\to n),
\label{t-unitarity}
\end{equation}
where the sum on the right-hand side is
taken over all allowed intermediate states $n$ having the  same total
4-momentum as  the initial state. In the following we restrict ourselves
to two-pion intermediate states, i.e. $n=\pi\pi$.

The 
amplitudes $T(\gamma\gamma\to\pi\pi)$ and  $T(\pi\pi\to\bar{N}N)$
are  constructed  by making use of available experimental 
information on two different reactions. For the amplitude
$T(\gamma\gamma\to\pi\pi)$ this is  the two-photon fusion reaction
$\gamma\gamma\to\pi\pi$. Since
there are no data on the reaction $\pi\pi\to\bar{N}N$,
the amplitude $T(\pi\pi \to \bar{N}N)$ is constructed in a dispersive approach
using the well known amplitudes of the pion-nucleon scattering reaction
$\pi N \to N \pi$. At $t>m^2_\pi$ unitarity shows that the phases of the
amplitudes 
$T(\gamma\gamma\to\pi\pi)$ and  $T(\pi\pi\to\bar{N}N)$
are the same and equal to the phases of pion-pion scattering, $\delta^J_I$, 
which are known from the data on the reaction $\pi p\to p\pi\pi$.

The phase-dependent factor entering into  the amplitude of a 
narrow resonance is  described  by a 
Breit-Wigner curve, whereas for the present case of a very broad resonance 
as given by the functions $\delta^J_I(t)$, a generalized version of the
Breit-Wigner curve has to be used.
This generalized  phase-dependent  factor is given
by the Omn\`es \cite{omnes58} function
$\Omega^J_I(t)$  defined through
\begin{equation}
\Omega^J_{I}(t)={\rm exp}{\left[
\frac{t}{\pi}\int^\infty_{4m^2_\pi}dt'\frac{\delta^J_{I}(t')} 
{t'(t'-t-i 0)}\right]}\equiv e^{i\delta^J_{I}(t)}
{\rm exp}{\left[\frac{t}{\pi}{\cal P}\int^\infty_{4m^2_\pi}
\frac{\delta^J_{I}(t')dt'}{t'(t'-t)} \right]}.
\label{Omnesfunct}
\end{equation}

For the discussion of the properties of the $T(\pi\pi\to \bar{N}N)$
amplitude in the unphysical region  $4m^2_\pi\leq t \leq 4M^4$  
we use  the backward amplitude
$F^{(+)}(t)$ discussed  by Bohannon \cite{bohannon76}. For the pion
scattering process  the quantity $t$ is negative, whereas for positive
$t$ the analytic continuation of $F^{(+)}(t)$ describes the
$N\bar{N}\to \pi\pi$ annihilation process for the case that the
helicities of the nucleon and the antinucleon are the same, 
$\lambda = \bar{\lambda}$. This provides us with a
tool to construct the $N\bar{N}\to \pi\pi$ amplitude from the measured
$\pi N\to \pi N$ amplitude. Since the backward $\pi N\to N\pi$
scattering amplitude implies also backward $N \bar{N} \to \pi\pi$
annihilation, it can be shown that the first two terms of the expansion 
for $F^{(+)}(t)$
at positive $t$ are
\begin{equation}
F^{(+)}(t)= \frac{16 \pi}{m(4m^2-t)}f^{0}_+(t) -\frac{5\pi(t-4m^2_\pi)}
{m} f^2_+(t).
\label{bohannon}
\end{equation}
The amplitudes $f^J_+(t)$ are the partial wave amplitudes
introduced  by Frazer and Fulco 
\cite{frazer60}, where $J$ denotes the angular momentum of the
$\pi\pi$ intermediate state. The $+$ sign denotes that the
two  helicities $\lambda$ and $\bar{\lambda}$ of  $N$ and $\bar{N}$,
respectively, are the same, $\lambda = \bar{\lambda}$.

The construction of both amplitudes, $T(\gamma\gamma\to\pi\pi)$ and
$T(\pi\pi \to \bar{N}N)$, in connection with the scalar-isoscalar $t$-channel
of Compton scattering has first been described and worked out in some detail
by K{\"o}berle \cite{koeberle68} and later discussed  by several 
authors, of whom we wish to cite 
\cite{morgan88,holstein94,drechsel99,drechsel03,levchuk04}. 
For a very broad resonance the appropriate ansatz reads
\begin{equation}
F^J_{I\lambda}(t)=\Omega^J_I(t)   P^J_{I\lambda}(t).
\label{gg-pipi-ansatz}
\end{equation}
where $P^J_{I\lambda}(t)$ is a real amplitude in the 
$\gamma\gamma\to\pi\pi$ physical region  and 
$\Omega^J_I(t)$ the  phase-dependent Omn\'es function discussed above.
In (\ref{gg-pipi-ansatz}) $I$ is the isospin of the
transition, $J$ the angular momentum and 
$\lambda\equiv \Lambda^t_{\gamma\gamma}$ 
the helicity
difference of the two photons. In the present case we 
have $\lambda\equiv \Lambda^t_{\gamma\gamma}\equiv 0$, 
so that we can omit the index
$\lambda$ without loss of generality. 
The amplitudes $F^J_I(t)$ have to be constructed such that they
have the correct low-energy properties,
 reproduce the 
cross section of the photon fusion reaction $\gamma\gamma\to \pi\pi$ and 
incorporate the phases $\delta^J_I(t)$
\cite{colangelo01,kaloshin94,pennington97,hyams73,froggatt77}.
For  $J=0$  the following form of Eq. 
(\ref{gg-pipi-ansatz}) has been obtained \cite{levchuk04}:
\begin{eqnarray}
F^0_{I}(t)&=&\Omega^0_I(t)\Big\{\left[F^{B,0}_{I}(t)+ \Delta
 F^0_{I}(t)\right] 
{\rm Re}\frac{1}{\Omega^0_I(t)} -\frac{t^2}{\pi}\Big[ {\cal P} \int^\infty
_{4 m^2_\pi}[ F^{B,0}_{I}(t')+\Delta
 F^0_{I}(t')] {\rm Im}\frac{1}{\Omega^0_I(t')}
\frac{dt'}{t'^2(t'-t)}\nonumber\\
&+& A^0_I + t B^0_I\Big]\Big\}.
\label{modified-s-wave}
\end{eqnarray}
The expression in Eq. (\ref{modified-s-wave}) is given for  the $s$-wave
amplitude where contributions going beyond the Born approximation
have been taken into account.
The leading term in (\ref{modified-s-wave}) is a superposition of a
Born term, $F^{B,0}_{I}(t)$, and a pion-structure dependent correction,
$\Delta F^{0}_{I}(t)$.
Since the reactions $\gamma\gamma\to \pi^+\pi^-,\pi^-\pi^+,\pi^0\pi^0$
show up with two components having  isospin $I=2$ and one component having
$I=0$ we have to take into account these two isospins with the appropriate 
weights (for details see \cite{schumacher05}).

The amplitudes of interest for the prediction of 
$(\alpha-\beta)^t$
are the $S$-wave amplitude $F^0_{0}(t)$ and the $D$-wave amplitude
$F^2_{0}(t)$ with the $D$-wave amplitude  leading to only a small
correction. It, therefore, is appropriate to restrict the discussion
mainly to the amplitude $F^0_{0}(t)$. The essential
property  of this  amplitude is  provided  by the Omn\`es function
which introduces a zero crossing of the amplitude at about 570 MeV.
This zero crossing may  be considered as a manifestation of 
the $\sigma$ meson which shows up through the phase-shift 
$\delta^0_0(t)$ of the correlated $\pi\pi$ pair.

If we restrict ourselves in the calculation
of the $t$-channel absorptive part to intermediate states with two
pions with angular momentum $J\leq 2$, the sum rule (\ref{SR})
takes the convenient form for calculations \cite{bernabeu77}:
\begin{eqnarray}
(\alpha-\beta)^t&=& 
 \frac{1}{16 \pi^2}\int^\infty_{4 m^2_\pi}\frac{dt}{t^2}\frac{16}{4m^2-t}
\left(\frac{t-4m^2_\pi}{t}\right)^{1/2}\Big[f^0_+(t)
  F^{0*}_{0}(t)\nonumber\\
&&-\left(m^2-\frac{t}{4}\right)\left(\frac{t}{4}-m^2_\pi\right)
f^2_+(t) F^{2*}_{0}(t)\Big],\label{BackSR}
\end{eqnarray}
where
$f^{(0,2)}_+(t)$ and $F^{(0,2)}_0(t)$ are the partial-wave helicity
amplitudes of the processes $N\bar{N}\to \pi\pi$ and 
$\pi\pi\to \gamma\gamma$ with angular momentum $J=0$ and $2$,
respectively, and isospin $I=0$.

\subsection{Test of the BEFT sum rule}

Though being the first who published the BEFT sum rule in its presently
accepted form, Bernabeu and Tarrach \cite{bernabeu77}
were not aware of the appropriate amplitudes  to calculate this sum rule 
numerically.   Also the first calculation of Guiasu and Radescu
\cite{guiasu76} remained incomplete because the amplitudes were used in the
form of their Born approximation  and, as the major drawback,  the correlation
of pions was not taken into account.  
Table \ref{BEFtable} summarizes those results
\begin{table}[h]
\caption{Numerical evaluation of the BEFT  sum rule, with corrections a) 
and b) supplemented by the present author. The unit is $10^{-4}{\rm fm}^3$.}
\vspace{5mm}
\begin{center}
\begin{tabular}{|l|l|l|l|}
\hline
$(\alpha_p-\beta_p)^s$&$(\alpha_p-\beta_p)^t$&$(\alpha_p-\beta_p)$
&authors\\
\hline
$-$4.92&+9.28$^{a)}$&+4.36&Guiasu, Radescu,1978 \cite{guiasu78}\\
$-$4  & +10.4$^{b)}$ & +6.4 & Budnev, Karnakov, 1979 \cite{budnev79}\\
$-$5.42&+8.6&$+(3.2^{+2.4}_{-3.6})$&{Holstein,Nathan,1994}\cite{holstein94}\\
$-$5.56&+16.46&+(10.7$\pm$0.2)$^{c)}$&{Drechsel,Pasquini,
Vanderhaeghen,2003}\cite{drechsel03}\\
$-(5.0\pm 1.0)$&$+(14.0\pm 2.0)$&$+(9.0\pm 2.2)$&Levchuk, 2004 \cite{levchuk04}\\
\hline
$-4.96$&$+15.0^{d)}$&$+10.0$&present prediction\\
&&$+10.1\pm 0.8$&experiment\\
\hline
\end{tabular}
\end{center}
a) corrected for the $D$-wave 
contribution $(-1.7)$ included. In an
earlier work Guiasu and Radescu \cite{guiasu76} used the Born
approximation without $\pi\pi$ correlation for both 
amplitudes $N\bar{N}\to \pi\pi$ and
$\pi\pi\to\gamma\gamma$ 
and arrived at $(\alpha-\beta)^t= +17.51$.\\
b) correction for the polarizability of the pion (+3.0) included.\\
c) best value from a range of results given by the authors
\cite{drechsel03}.\\
d) $\sigma$-pole contribution. 
\label{BEFtable}
\end{table}
of tests of the BEFT sum rule where at least the $\pi\pi$ correlation is taken
into account. In the early works of Guiasu and Radescu \cite{guiasu78}
and Budnev and Karnakov \cite{budnev79} some missing pieces in the results
were identified which are supplemented in Table \ref{BEFtable} on the basis
of the results given by Holstein and Nathan \cite{holstein94}.
In case of the Holstein and
Nathan result \cite{holstein94} we interpret the rather conservatively
estimated upper and lower bounds as errors.   In case of the Drechsel
et al. result \cite{drechsel03} we quote the $s$-channel and $t$-channel
contributions calculated at $\theta=180^\circ$. The total result
$(\alpha_p-\beta_p)$ is the best value for this quantity 
extracted by the authors from results obtained in the angular region
$140^\circ \leq \theta_{\rm lab} \leq 180^\circ$. 

As will be shown in the following the present prediction given in line 7 of
Table \ref{BEFtable} is much more precise than  the previous predictions
given in line 2--6. The reason is that the rather complicated evaluation 
of the BEFT sum rule is replaced by the straightforward 
evaluation of the 
well founded $\sigma$-meson pole.
Furthermore, it will be shown that the $\sigma$-meson pole has a transparent
interpretation in terms of chiral symmetry breaking. The $(\pibf,\sigma)$
quartet may be considered as part of the mesonic structure of the constituent
quark. Of these mesons the $(\pi^0,\sigma)$ doublet has the capability of
interacting  with two photons and, therefore, to contribute to nucleon Compton
scattering and the polarizabilities.

\subsection{Comparison of the BEFT sum rule and the $\sigma$ meson pole}

The   $\pi^0$
singularity enters into the invariant amplitude $A_2$ and the $\sigma$ 
singularity into the invariant amplitude $A_1$ \cite{lvov97}.
Starting  this section with the discussion of the properties of the 
$\pi^0$-meson pole we may write down
\begin{equation}
{\rm Im}_t A^{\pi^0}_2(t)=\pi {\cal M}(\pi^0\to\gamma\gamma)g_{\pi NN}
\delta(t-m^2_{\pi^0})\tau_3
\label{unitarity}
\end{equation}
where ${\cal M}(\pi^0\to\gamma\gamma)$ is the transition matrix element and 
$g_{\pi NN}$ the pion nucleon coupling constant and $m_{\pi^0}$ the
$\pi^0$ mass.
The dispersion relation for the $\pi^0$-pole contribution is given by
\begin{equation}
A_2^{\pi^0}(t)=\frac{1}{\pi}\int^\infty_{t_0} {\rm Im}_t 
A_2(t')
\frac{d t'}{t'-t-i0}
\label{int}
\end{equation}
with the solution for a point like singularity \cite{lvov99}
\begin{equation}
A^{\pi^0}_2(t)=\frac{ {\cal M}(\pi^0 \to \gamma\gamma)\, g_{\pi NN} }
{t-m^2_{\pi^0}}\taubf_3.
\label{pipole}
\end{equation}
Though the foregoing  few steps of arguments are very straightforward the
interpretation of the result given in Eq. (\ref{pipole}) needs some 
comments referring to the kinematics of the two photons. In two-photon
decay and two-photon fusion reactions the two photons
move in opposite directions.
Translating this into a Compton scattering process we arrive at the conclusion
that the scattering angle is $\theta=\pi$. Smaller scattering angles apparently
require a kinematical correction which will be considered later.
In addition there are contributions from the $\eta$ and $\eta'$ mesons
so that the pseudoscalar $t$-channel amplitude for $\theta=\pi$ becomes 
\begin{equation}
A^{{\pi^0}+\eta+\eta'}_2(t)=\frac{g_{{\pi^0} NN}{\cal M}(\pi^0\to\gamma\gamma)}
{t-m^2_{\pi^0}}\tau_3+\frac{g_{{\eta} NN}{\cal M}(\eta\to\gamma\gamma)}
{t-m^2_{\eta}}+\frac{g_{{\eta'} NN}{\cal M}(\eta'\to\gamma\gamma)}
{t-m^2_{\eta'}},
\label{pseudoscalarPole-1}
\end{equation}
where the quantities $g_{\pi^0 NN}$, etc. are the meson-nucleon coupling
constants and the quantities ${\cal M}(\pi^0 \to\gamma\gamma)$, etc. the 
two-photon decay amplitudes. 

This leads us to the following
consideration for the scalar counterpart of the pseudoscalar $t$-channel
contribution.  The BMFT sum rule calculates $(\alpha-\beta)^t$ 
from the reaction chain
\begin{equation}
\gamma\gamma\to\sigma\to\pi\pi\to\sigma\to N\bar{N}
\label{chain}
\end{equation}
with small additional contributions from the $f_0(980)$ and $a_0(980)$
mesons. These  scalar mesons are known to exhaust the 
$\gamma\gamma\to$meson-meson channel below the $N\bar{N}$ threshold. 
\begin{figure}[h]
\centering\includegraphics[width=0.5\linewidth]{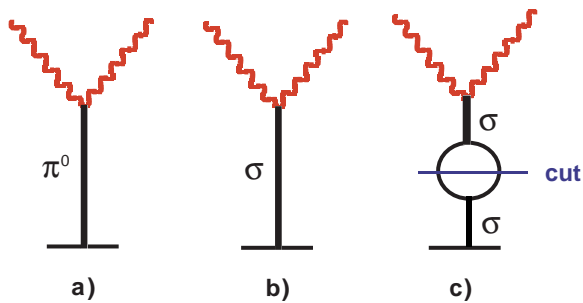}
\caption{$\pi^0$ and $\sigma$ pole graphs.}
\label{pi-sigma-pole}
\end{figure}
Now we notice that Eq. (\ref{chain}) describes a computational method
but not the $t$-channel Compton 
scattering process, which
is described through
\begin{equation}
\gamma\gamma\to\sigma\to N\bar{N}.
\label{reaction}
\end{equation}
The reason is that in the Compton scattering process only one intermediate
state is possible, i.e. the creation of a $\sigma$ meson as part of the
constituent-quark structure, whereas two or more intermediate states following
each 
other are forbidden.
This difference between computational method and  $t$-channel
Compton scattering is illustrated 
by the graphs in Figure \ref{pi-sigma-pole}. Graph  b) describes 
the $t$-channel scattering process and graph c) the computational
method where an intermediate pion loop is introduced so that the BEFT
sum rule can be evaluated. On the basis of this consideration we can write
down the scalar analog of Eq. (\ref{pseudoscalarPole-1}) in the form
\begin{equation}
A^{{\sigma}+f_0+a_0}_1(t)=\frac{g_{{\sigma} NN}{\cal M}(\sigma\to\gamma\gamma)}
{t-m^2_{\sigma}}+\frac{g_{{f_0} NN}{\cal M}(f_0\to\gamma\gamma)}
{t-m^2_{f_0}}+\frac{g_{{a_0} NN}{\cal M}(a_0\to\gamma\gamma)}
{t-m^2_{a_0}}\tau_3.
\label{scalarPole}
\end{equation}
At this point the mass  $m_\sigma$ is a parameter and 
it has to be clarified how the mass $m_\sigma$
has to be interpreted. This will be done in the next subsection.

\subsection{Structure of the $\sigma$ meson}

We restrict the present outline to the main argument and refer to 
\cite{tornqvist95,boglione 02,schumacher11b} for further details. The $\sigma$
meson is a strongly decaying particle which may be described by a propagator
in its most general form
\begin{equation}
P(s)=\frac{1}{m^2_b+\Pi(s)-s}
\label{prop1}
\end{equation}
where $m_b$ is the ``bare'' mass\footnote{The term ``bare'' mass is used to
  describe the mass of a particle where only the $\gamma\gamma$ 
decay channel is open and correspondingly
the meson-meson channels are closed} and $\Pi(s)$ 
a ``vacuum polarization function''
\cite{tornqvist95,boglione 02}. The imaginary part of $\Pi(s)$ may be
obtained from unitarity considerations applied to the meson-meson decay
  channels. Since the ``vacuum polarization
function'', $\Pi(s)$, is  analytic, its real part can be 
deduced from
the imaginary part by making use of a dispersion relation. In a further
step we may write the propagator in the form
\begin{equation}
P(s)=\frac{1}{m^2(s)-s-i\,m_{\rm BW}\Gamma_{\rm tot}(s)},
\label{prop2}
\end{equation}
having identified 
\begin{equation}
\Gamma_{\rm tot}=-\frac{{\rm Im}\Pi(s)}{m_{\rm BW}}
\label{prop3}
\end{equation}
and 
\begin{equation}
m^2(s)=m^2_b+{\rm Re}(s)
\label{prop4}
\end{equation}
where $m^2(s)$ is the running mass square and $m_{\rm BW}$ the Breit-Wigner
mass given by
\begin{equation}
m^2_{\rm BW}=m^2_b+{\rm Re}\Pi(m^2_{\rm BW}).
\label{prop4}
\end{equation} 
Up to this point the variable $s$ is a real quantity. But it is possible
to shift this variable $s$ into the complex plane such that the denominator
in Eq. (\ref{prop1}) is equal to zero. Then the propagator may be written in
the form
\begin{equation}
P(s)=\frac{1}{s_R-s}
\label{prop5}
\end{equation}
corresponding to a pole on the second  Riemann sheet at 
$s_R=(M_R-\frac12\,i\,\Gamma_R)^2$.
For the $\sigma$-meson the ``bare'' mass$^2$ is
$m_\sigma=666$ MeV (see \cite{schumacher10} and the next subsection)
whereas for the quantities
$M_\sigma$ and $\Gamma_{\sigma}$ the results $M_\sigma=441^{+16}_{-8}$ MeV
and $\Gamma_\sigma=544^{+18}_{-25}$ \cite{caprini06} have been obtained.
Further details may be found in \cite{schumacher10,scadron08}.

In case of Compton scattering the two-photon fusion 
reaction $\gamma\gamma\to \sigma \to N\bar{N}$ has to be 
considered 
instead of the reaction $\gamma\gamma\to\sigma \to \pi\pi$. This leads to the
consequence that $\Pi(s)\equiv 0$, because obviously there is no open
meson-meson decay channel.
Therefore, the propagator given in Eq.
(\ref{prop1}) has the form
\begin{equation}
P(s)=\frac{1}{m^2_b -s}.
\label{propagator1}
\end{equation}
For the Compton scattering amplitudes the pole described in 
Eq. (\ref{propagator1})
corresponds to a pole on the positive $t$-axis at $t_0=m^2_b$.
This leads to the scalar pole terms given in Eq. (\ref{scalarPole})
by identifying $m_b$ with $m_\sigma$ and the other masses, respectively.

\subsection{The quark level linear $\sigma$ model (QLL$\sigma$M)}

Now we come to the fundamental problem how the $\sigma$ meson 
and the other scalar mesons enter into the structure of the nucleon.
First we notice that quite obviously the constituent quarks couple to all
mesons having a nonzero meson-quark coupling constant. This statement is
nontrivial because there are approaches where only Goldstone bosons are taken 
into consideration. Second we restrict the discussion to $SU(2)$ and refer to
a generalization to $SU(3)$ given in \cite{schumacher11b}. Then we arrive at
the description of chiral symmetry breaking given in the following.

The QLL$\sigma$M \cite{delbourgo95}
combines the linear $\sigma$ model (\ref{ql1}) with two
versions of the Nambu--Jona-Lasinio model \cite{nambu61}, 
the four-fermion versions of
Eq. (\ref{ql2}) and bosonized versions of Eq. (\ref{ql3})
\begin{eqnarray}
&&{\cal L}_\sigma=\frac12\partial_\mu\pibf\cdot\partial^\mu
\pibf + \frac12 \partial_\mu\sigma\partial^\mu\sigma 
+\frac{\mu^2}{2}(
\sigma^2+\pibf^2)-\frac{\lambda}{4}(\sigma^2+\pibf^2)^2+f_\pi m^2_\pi 
\sigma, \label{ql1}\\
&&{\cal L}_{\rm NJL}=\bar{\psi}(i\fpartial-m_0) \psi
+
\frac{G}{2}[(\bar{\psi}\psi)^2+(\bar{\psi}i\gamma_5\vectau\psi)^2],\label{ql2}
\\
&&{\cal L'}_ {\rm
  NJL}=\bar{\psi}i\fpartial\psi-g\bar{\psi}(\sigma+i\gamma_ 5
\vectau\cdot\pibf)\psi-\frac12\delta\mu^2(\sigma^2+\pibf^2)+\frac{gm_0}{G}
\sigma, \label{ql3}\\
&&G=g^2/\delta\mu^2, \quad 
\delta\mu^2=(m^{\rm cl}_\sigma)^2,\quad 
G=\lambda/(\sqrt{2}m^{\rm cl}_\sigma)^2, \quad g=\sqrt{\lambda/2}, \label{ql4}
\end{eqnarray}
where $m^{\rm cl}_\sigma$ denotes the mass of the $\sigma$ meson in the chiral
limit (cl). Eq. (\ref{ql1}) describes spontaneous chiral symmetry breaking
in terms of fields
where a mexican hat potential is introduced, parameterized by the mass
parameter 
$\mu^2>0$ and the self-coupling parameter $\lambda>0$. Explicit symmetry
breaking is taken into account by the last term which vanishes in the chiral
limit $m_\pi \to 0$. In contrast to this Eqs. (\ref{ql2}) and (\ref{ql3})
describe dynamical chiral symmetry breaking in terms of $u$ and $d$
constituent quarks
as the relevant fermions. In Eq. (\ref{ql2}) the interaction between
the fermions is parameterized by the four-fermion interaction constant $G$
and explicit symmetry breaking by the average current-quark mass $m_0$.
Eq. (\ref{ql3}) differs from  (\ref{ql2}) by the fact that the interaction
is described by the exchange of bosons. This leads to the occurrence of a mass
counter term parameterized by $\delta\mu^2$. Eq. (\ref{ql4}) shows how
these different parameters are related to each other.

In more detail, the bosonization starts with the following ansatz \cite{vogl91}
\begin{eqnarray}
&&\sigma=-\frac{G}{g}\bar{\psi}\psi, \label{bos1}\\
&&\pibf=-\frac{G}{g}\bar{\psi}\,i\,\gamma_5\taubf\,\psi. \label{bos2}
\end{eqnarray}
This ansatz leads to the relation
\begin{equation}
-\frac{g^2}{2G}(\sigma^2+\pibf^2)-\bar{\psi}\,g\,(\sigma+i\,\gamma_5\taubf)\psi
=\frac{G}{2}\left[(\bar{\psi}\psi)^2+(\bar{\psi}i\,
\gamma_5\,\taubf\,\psi)^2\right]
\label{bos3}
\end{equation}
and therefore converts
\begin{equation}
{\cal L}=\bar{\psi}(i\,/\!\!\!\partial-m_0)\psi+\frac{G}{2}
\left[(\bar{\psi}\psi)^2+\bar{\psi}i\,\gamma_5\,\taubf\,\psi)^2\right]
\label{bos4}
\end{equation}
into
\begin{equation}
{\cal L}=\bar{\psi}(i\,/\!\!\!\partial-g(\sigma+
\pibf\cdot\taubf\,\gamma_5))\psi
-\frac{g^2}{2G}(\sigma^2+\pibf^2)+\frac{gm_0}{G} \sigma.
\label{bos5}
\end{equation}

The use of two versions of the NJL model has the advantage that for 
many applications the four-fermion version is more convenient whereas
the bosonized version describes the interaction of the constituent quark
with the QCD vacuum  through the exchange of the $(\pibf,\sigma)$ meson quartet
which we consider as the true description of the physical process.

Using diagrammatic techniques the following  equations may be found 
\cite{klevansky92,hatsuda94}  for the non-strange $(\pibf,\sigma)$ sector
\begin{eqnarray}
&&M^*=m_0+ 8\, i\, G N_c \int^{\Lambda}\frac{d^4 p}{(2\pi)^4}
\frac{M^*}{p^2-M^{*2}},\quad M=-\frac{8\,iN_cg^2}{(m^{\rm cl}_\sigma)^2}
\int\frac{d^4p}{(2\pi)^4}\frac{M}{p^2-M^2},
\label{gapdiagram}\\
&&f^2_\pi = -4\,i\,N_cM^{*2} \int^{\Lambda}\frac{d^4p}{
(2\pi)^4}\frac{1}{(p^2-M^{*2})^2},\quad f^{\rm cl}_\pi=-4iN_cgM\int\frac{d^4p}{(2\pi)^4}
\frac{1}{(p^2-M^2)^2}.
\label{fpiexpress}
\end{eqnarray}
The expression given on the l.h.s. of (\ref{gapdiagram}) is the 
gap equation with $M^*$
being the mass of the constituent quark with the contribution 
$m_0$ of the
current quarks included. The r.h.s.
shows the gap equation 
for the nonstrange $(n\bar{n})$ constituent-quark mass $M$ in the chiral limit.
The l.h.s. of Eq.
(\ref{fpiexpress}) represents  the pion decay constant and the r.h.s.
the same quantity in the chiral limit. 
 For further details we refer to 
\cite{schumacher06,delbourgo95}.

Making use of dimensional regularization the
Delbourgo-Scadron \cite{delbourgo95} relation 
\begin{equation}
M=\frac{2\pi}{\sqrt{N_c}}f^{\rm cl}_\pi, \quad N_c=3
\label{sigmamass}
\end{equation}
may be obtained from 
the r.h.s of Eqs. (\ref{gapdiagram}) and (\ref{fpiexpress}).
This important relation shows that the mass of the constituent quark 
in the chiral limit and the pion decay constant in the chiral limit are
proportional to each other. This relation is valid independent of the
flavor content of the constituent quark, e.g. also for a constituent quark
where the  $d$-quark is replaced by a $s$-quark. Furthermore, it has been
shown \cite{delbourgo98,delbourgo02} that (\ref{sigmamass}) is  
valid independent of the
regularization scheme.

An update for the mass prediction for the $\sigma$ mesons is obtained
by using the more recent result for the pion decay constant as given in
\cite{PDG}.
This leads to the neutral-pion decay constant
\begin{equation}
f_\pi=(92.21\pm 0.15)\,\, {\rm MeV}
\label{new1}
\end{equation}
and via a once-subtracted dispersion relation \cite{coon81}
\begin{equation}
1-\frac{f^{\rm cl}_\pi}{f_\pi}=\frac{m^2_\pi}{8\pi^2f^2_\pi}= 2.8\%
\label{new2}
\end{equation}
to
\begin{equation}
f^{\rm cl}_\pi=f_\pi(1-0.028)=89.63 \,\,{\rm MeV}.
\label{new3}
\end{equation}
Then the update of the mass of the $\sigma$ meson is
\begin{equation}
m^{\rm cl}_\sigma=650.3\,\,{\rm MeV}
\label{new4}
\end{equation}
and
\begin{equation}
m_\sigma=[(m^{\rm cl}_\sigma)^2+{\hat m}^2_\pi]^{1/2}= 664 \,\, {\rm MeV}
\label{new5}
\end{equation}
which agrees with the standard value \cite{schumacher06}
$m_\sigma=666$ MeV within  0.3\%.

It is interesting to note that  the QLL$\sigma$M for $N_c=3$ nonperturbatively
predicts the parameters of the potential entering into the linear $\sigma$
model to be
 $\lambda=\frac{8\pi^2}{3}=26.3$ and 
$\mu=\sqrt{\frac23}2\pi f^{\rm cl}_\pi=459.8$
MeV, whereas these quantities remain undetermined when only
spontaneous 
symmetry breaking as given in (\ref{ql1}) is considered.

For sake of completeness we wish to mention that in addition to the aspects 
of the QLL$\sigma$M as outlined above we may 
write the  SU(2) interaction part of the QLL$\sigma$M Lagrangian density 
in the form \cite{delbourgo95,beveren02}
\begin{equation}
{\cal L}_{L \sigma M}^{\rm int}=g\bar{\psi}(\sigma+i\gamma_5\taubf\cdot\pibf)
\psi+g'\sigma(\sigma^2+\pibf^2)-\frac{\lambda}{4}(\sigma^2+\pibf^2)^2.
\label{scad}
\end{equation}
Here  $\sigma$ denotes the $\sigma$ field after the shift of its 
origin to the
vacuum expectation value $f_\pi$ has been carried out. The fermion 
fields refer to 
constituent quarks as described in
\cite{delbourgo95,beveren02}. Eq. (\ref{scad}) allows to generate expression 
for the $\pi-\sigma-\pi$ and $\sigma-\sigma-\sigma$ coupling constants
which, however, are not of importance for the present considerations.

Of the different options to describe chiral symmetry breaking we adopt the
bosonized NJL model as the one which describes the $\sigma$ meson 
as part of  constituent-quark structure in the most transparent way. The linear
$\sigma$ model is only discussed for sake of completeness. Simultaneously a
definite prediction of the $\sigma$ meson mass is obtained. The prediction
of the transition amplitude ${\cal M}(\sigma\to\gamma\gamma)$ requires
the knowledge of the quark-structure of the $\sigma$ meson where we have the 
option of a $q\bar{q}$ structure and a tetraquark structure. As has been shown
in detail in \cite{schumacher11b} these two options do not make a difference 
in the present case because the two-photon interaction takes place 
with the $q\bar{q}$ structure
in any case. This $q\bar{q}$ structure either serves as the intermediate state
in nucleon Compton scattering or as a doorway state in case of the tetraquark
structure \cite{schumacher11b} in the reaction chain 
$\gamma\gamma\to (u\bar{u}+d\bar{d})/\sqrt{2} \to u\bar{u}d\bar{d}\to\pi\pi$.
Therefore, the $q\bar{q}$ structure may be
used  for the 
prediction of  ${\cal M}(\sigma\to\gamma\gamma)$.

The results obtained in the present  subsection may be compared with a
more general treatment of the topic in ``Constituent-quark masses and the
electroweak standard model'' \cite{scadron06}.

\subsection{The QLL$\sigma$M  and polarizabilities}

As before we start with pseudoscalar  mesons and the corresponding
 spin-polarizability and thereafter turn to  scalar  mesons and 
the corresponding polarizability difference  $(\alpha-\beta)^t$.

 The main part of the $t$-channel component
of the backward spinpolarizability is given by
\begin{eqnarray}
&&|\pi^0\rangle=\frac{1}{\sqrt{2}}
(-|u\bar{u}\rangle+|d\bar{d}\rangle),\quad
{\cal M}(\pi^0\to\gamma\gamma)=\frac{\alpha_{em}N_c}{\pi f_\pi}
\left[-\left(\frac23\right)^2
+\left(\frac{-1}{3}\right)^2\right], \label{sp1}\\
&&{\gamma_\pi^t}_{(p,n)}=\frac{g_{\pi^0 NN} \quad
{\cal M}(\pi^0\to\gamma\gamma)}
{2\pi m^2_{\pi^0} m}\tau_3= -46.7\,\tau_3\,\,\, {10}^{-4}{\rm fm}^4.
\label{sp2}
\end{eqnarray}
Analogously, we obtain for the main $t$-channel parts of the electric
($\alpha$) and magnetic ($\beta$) polarizabilities the relations 
\begin{eqnarray}
&&|\sigma\rangle=\frac{1}{\sqrt{2}}(|u\bar{u}\rangle
+|d\bar{d}\rangle), \quad
{\cal M}(\sigma\to\gamma\gamma)=\frac{\alpha_{em}N_c}{\pi f_\pi}
\left[\left(\frac23\right)^2
+\left(\frac{-1}{3}\right)^2\right],\label{sp3}\\
&&(\alpha-\beta)^t_{p,n}=\frac{g_{\sigma NN}{\cal  M}(\sigma\to\gamma\gamma)}
{2\pi m^2_\sigma}= 15.2\,\,\, {10}^{-4}{\rm fm}^3,\quad
(\alpha+\beta)^t_{p,n}=0,\\
&&\alpha^t_{p,n}=+7.6\,\,\, {10}^{-4}{\rm fm}^3,\,
\beta^t_{p,n}=-7.6\,\, {10}^{-4}{\rm fm}^3\label{sp4},
\end{eqnarray}
where use is made of $g_{\pi NN}=g_{\sigma NN}=13.169\pm 0.057$ 
\cite{bugg04}
and
$m_\sigma=664$ MeV as predicted by the QLL$\sigma$M. The sign convention
used in the $q\bar{q}$ structure of the $\pi^0$ meson follows from 
\cite{close79}. It has the advantage of correctly predicting the sign of the
$\pi^0$-pole contribution.
These main contributions to the polarizabilities $\alpha$, $\beta$ and
$\gamma_\pi$ have to be supplemented by the $s$-channel components 
and by the small components due to the scalar mesons $f_0(980)$ and 
$a_0(980)$ in case of the polarizabilities $\alpha$ and $\beta$ and 
due to the pseudoscalar meson $\eta$ and $\eta'$ in case of $\gamma_\pi$.

The appropriate tool for the prediction of  the polarizabilities 
$\alpha$ and $\beta$
is to simultaneously apply
the forward-angle sum rule for $(\alpha+\beta)$ and the backward-angle sum
rule for $(\alpha-\beta)$. This leads to the  relations
derived in  \cite{schumacher07a,schumacher09} and shown in subsection 2.5.
From the expression for the scalar $t$-channel amplitude
given in (\ref{scalarPole}) we obtain
\begin{equation}
(\alpha-\beta)^t=\frac{g_{\sigma NN}{\cal M}(\sigma\to\gamma\gamma)}{2\pi m^2_\sigma}
+\frac{g_{f_0 NN}{\cal M}(f_0 \to\gamma\gamma)}{2\pi m^2_{f_0}}
+\frac{g_{a_0 NN}{\cal M}(a_0\to\gamma\gamma)}{2\pi m^2_{a_0}}\tau_3,
\label{pol8}
\end{equation}
and with $(\alpha+\beta)^t=0$
\begin{eqnarray}
&&\alpha^t=\frac12 (\alpha-\beta)^t,\label{alphat}\\
&&\beta^t=-\frac12 (\alpha-\beta)^t. \label{betat}
\end{eqnarray}.

The $t$-channel part of the backward spin-polarizability  
is given by \cite{schumacher07b,lvov99}
\begin{equation} 
\gamma^t_\pi=\frac{1}{2\pi m}\left[\frac{g_{\pi NN}
{\cal M}(\pi^0\to\gamma\gamma)}
{ m^2_{\pi^0}}\tau_3
+\frac{g_{\eta NN}{\cal M}(\eta \to\gamma\gamma)}{ m^2_{\eta}}
+\frac{g_{\eta' NN}{\cal M}(\eta'\to\gamma\gamma)}{
  m^2_{\eta'}}\right].\label{pol22}
\end{equation}

The numerical evaluation
of these  contributions has been described in detail in previous papers
\cite{schumacher07b,schumacher09}. Here we 
give a summary of the final predictions and the experimental values to compare
with in Table \ref{tab3} and \ref{tab4}. 
\begin{table}[h]
\caption{Backward spinpolarizabilities for the proton and the neutron
\cite{schumacher11a} and Table \ref{polresults1}} 
\begin{center}
\begin{tabular}{l|ll}
spin polarizabilities & $\gamma^{(p)}_\pi$ & $\gamma^{(n)}_\pi$\\
\hline
$\pi^0$ pole & -46.7 & +46.7\\
$\eta$ pole & +1.2 & +1.2\\
$\eta'$ pole & +0.4 & +0.4\\
\hline
const. quark structure&$-$45.1 &+48.3\\
nucleon structure & +8.5 & +10.0 \\
\hline
total predicted& $-$36.6 & +58.3\\
exp. result & 
$-$($36.4\pm 1.5$)& 
+($58.6\pm 4.0$)\\
\hline
&unit 10$^{-4}$ fm$^4$
\end{tabular} \label{tab3}
\end{center}
\end{table}
\begin{table}[h]
\caption{Polarizabilities for the proton and the neutron \cite{schumacher09}
and Table \ref{polresults1}} 
\begin{center}
\begin{tabular}{l|ll|ll}
 & $\alpha_p$ & $\beta_p$ & $\alpha_n$ & $\beta_n$\\
\hline
$\sigma$ pole & +7.6 & $-7.6$ & +7.6 &$-7.6$\\
$f_0$ pole & +0.3 & $-0.3$ & +0.3 & $-0.3$\\
$a_0$ pole & $-0.4$ & +0.4& +0.4 & $-0.4$\\
\hline
const. quark structure& +7.5 & $-7.5$
& +8.3 & $-8.3$\\
nucleon structure  & +4.5 & +9.4 & +5.1 & +10.1\\
\hline
total predicted& +12.0 & +1.9 & +13.4 & +1.8\\
exp. result  & 
+($12.0\pm 0.6$)
& +($1.9\mp 0.6$)& 
+($12.5\pm 1.7$)&
+($2.7\mp 1.8$) \\
\hline
&unit 10$^{-4}$ fm$^3$
\end{tabular}\label{tab4}
\end{center}
\end{table}
The most interesting feature  of the polarizabilities as given in Table
\ref{tab4} is the strong cancellation of the paramagnetic polarizabilities 
which is mainly due to the $P_{33}(1232)$ resonance by the diamagnetic
term which is solely due to the constituent-quark structure.

A further interesting conclusion concerns the  two-photon width of the
$\sigma$ mesons which can be calculated from the data given in 
(\ref{sp3}). Given the nonperturbative QLL$\sigma$M $\sigma$ meson mass of 664
MeV the amplitude for the  $\sigma\to\gamma\gamma$ decay for $N_c=3$ and
$f_\pi=92.21$ MeV is
\begin{equation}
{\cal M}(\sigma\to\gamma\gamma)=\frac{5\,\alpha_{\rm em}}{3\pi f_\pi}=0.042
\,\,\text{GeV}^{-1}
\label{num1}
\end{equation}
and the decay width
\begin{equation}
\Gamma_{\gamma\gamma}=\frac{m ^3_\sigma}{64\pi}
|{\cal M}(\sigma\to\gamma\gamma)|^2
=2.6\,\, \text{keV}.
\label{num2}
\end{equation}

From Table \ref{tab4} we obtain that the
predicted and the experimental electric polarizabilities $\alpha_p$ 
of the proton are in excellent agreement with each other. Furthermore, the
experimental quantity $\alpha_p$ has an experimental error of $\pm 5\%$
which may be used
to calculated the experimental error of the two-photon width  
$\Gamma_{\gamma\gamma}$ of the $\sigma$ mesons. This rather straightforward
consideration leads to $\Gamma_{\gamma\gamma}= 2.6\pm 0.3$ keV
\cite{schumacher10}.

\section{Compton scattering in a large angular range and at energies up to 1
  GeV}

In the foregoing we have restricted our considerations to the forward
and the backward directions. This is of advantage as long as we are interested
in theoretical aspects of nucleon Compton scattering and polarizabilities,
because transparent definitions of the polarizabilities are  possible
which may be used for theoretical predictions. Compton scattering 
experiments are only possible in an interval of intermediate angles
ranging from about $\theta_{\rm lab}=30^\circ$ to $150^\circ$. 
This means that the experimental determination of polarizabilities 
can only be obtained from data measured at these intermediate angles. In
addition 
to this the invariant amplitudes $A_i(s,t,u)$ (i=1--6) may be considered
as generalized polarizabilities containing valuable information about
interference 
effects between different phenomena contributing to the 
Compton scattering process. Of these the interference of scattering from the
$s$-channel and the $t$-channel is  of particular interest because it 
is related to degrees of freedom stemming from the photoexcitation
of the nucleon on the one hand and from photoexcitation of the mesonic
structure of the constituent quarks on the other. 
   
Compton scattering by the nucleon at energies up to 1.2 GeV 
has been studied
by  many groups in different laboratories 
over a long period of time. The final interpretation
of these experiments has been carried out by L'vov et al. in 1997
\cite{lvov97} where a precise method for the calculation of scattering
amplitudes is developed, based on non-subtracted fixed-$t$ dispersion
theory. This dispersion theory requires the introduction of asymptotic
or $t$-channel contributions, leading to the introduction of the
$\sigma$-meson pole amplitude in addition to the well known $\pi^0$-meson
pole amplitude. 
At this status of development the large-angle Compton scattering experiment
was carried at MAMI (Mainz) \cite{galler01,wolf01} which is already 
described in the introduction. This experiment precisely confirmed the
dispersion theory developed by L'vov et al. \cite{lvov97}. 

 In spite of
this great success fixed-$t$ dispersion theory  has  a disadvantage,
because the  integration paths for the $s$-channel dispersion relations
are to a large extent in the unphysical region. This is not of relevance for
the prediction of Compton differential cross sections but it
makes the interpretation of the amplitudes intransparent. 
\begin{figure}[h]
\centering\includegraphics[width=0.6\linewidth]{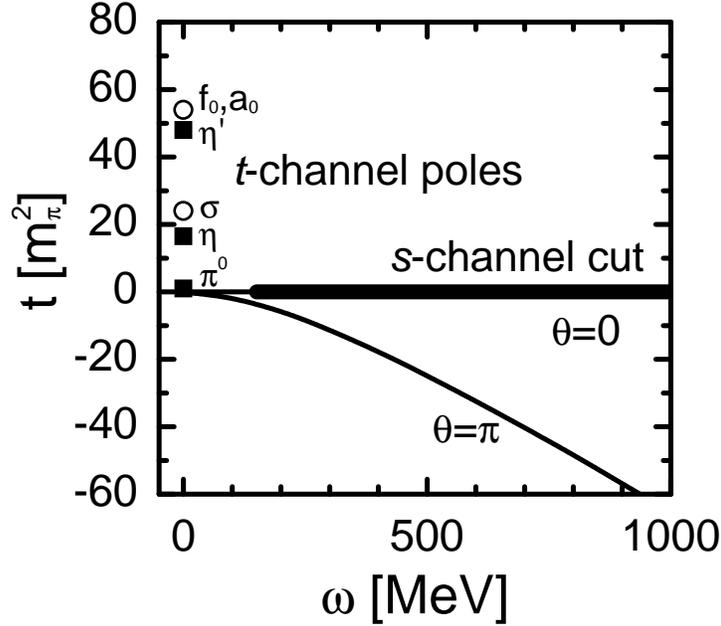}
\caption{Singularities  of the $s$- and the $t$-channel. Horizontal thick line:
$s$-channel cut representing the nucleon  (s-channel)
degrees of freedom, with $\omega$
being the energy of the incoming photon in the lab system. Squares and circles
on the 
vertical real $t$-axis: $t$-channel poles representing the constituent-quark
degrees of freedom. The physical range of nucleon Compton scattering extends
from the line at $\theta=0$ to the line at $\theta=\pi$.}
\label{s-t-plane}
\end{figure}
For illustration Figure \ref{s-t-plane} shows the physical range of 
Compton scattering and the location of the singularities of the Compton
scattering  amplitudes in the $s$-$t$-plane. At $\theta=0$ the integration path
of the dispersion integral coincides with the $s$-channel cut shown in 
Figure  \ref{s-t-plane}  which of course is located in the physical range.
However, for larger negative $t$ the integration paths enter the area
in the left part of the $s$-$t$-plan which is located outside the physical
plane.

Subtracted fixed-$t$ dispersion theory
\cite{drechsel03,drechsel99} has been applied and tested in the
$\Delta$-resonance region and below but there is no information how it
performs at higher energies.
In order to overcome the disadvantages inherent in fixed-$t$ dispersion
theories 
Drechsel et al. \cite{drechsel03} have proposed to use hyperbolic 
dispersion relations which are essentially equivalent to fixed-$\theta$
dispersion relations. 
As will be outlined in the following the integration paths of hyperbolic
dispersion relations  keep the angle $\theta_{lab}$ fixed whereas the angle
$\theta_{\rm cm}$ becomes  dependent on the photon energy
$\omega$. Fixed-$\theta $
or hyperbolic dispersion relations
are easy to apply at $\theta=180^\circ$ but become increasingly uncertain 
at smaller angles.   Especially, the evaluation of the $t$-channel
contribution based on a scalar-isoscalar $t$-channel cut breaks down at 
scattering angles of $\theta_{\rm lab}<101^\circ$.

This difficulty may be avoided by replacing
the $t$-channel cut by $t$-channel poles as provided by the QLL$\sigma$M.
Since this new type of approach has never before been investigated it is
appropriate to include a detailed description in  the present status report.

\subsection{Kinematics of Compton scattering}

The conservation of energy and momentum in nucleon Compton scattering
\begin{equation}
\gamma(k,\lambda) + N(p) \to \gamma'(k',\lambda')+N'(p')
\label{s-channel}
\end{equation} 
 is given by 
\begin{equation}
k+p=k'+p'\label{enermomen}
\end{equation}
where $k$ and $k'$ are the 4-momenta of the incoming and outgoing photon
and $p$ and $p'$ the 4-momenta of incoming and outgoing proton.
Mandelstam variables are introduced via
\begin{eqnarray}
&&s=(k+p)^2=(k'+p')^2, 
t=(k-k')^2=(p'-p)^2, 
u=(k-p')^2=(k'-p)^2,\\
&&s+t+u=2m^2. \label{constraint}
\end{eqnarray}
with $m$ being the nucleon mass.

In terms of the Mandelstam variables the 
scattering angle $\theta$ in the cm
system is given by
\begin{equation}
\sin^2\frac{\theta_{\rm cm}}{2}=-\frac{st}{(s-m^2)^2}.
\label{scatteringangle}
\end{equation}
The $t$-channel
corresponds to the fusion of two photons with four-momenta
$k_1$ and $k_2$ and helicities $\lambda_1$ and $\lambda_2$ to form
a $t$-channel intermediate state $|t\rangle$ from which -- in  a
second step -- a proton-antiproton pair is created.  
The corresponding reaction may be formulated in the 
form
\begin{equation}
\gamma(k_1,\lambda_1) + \gamma(k_2,\lambda_2)  \to  \bar{N}(p_1)  +N(p_2).
\label{t-channel}
\end{equation}
Since for Compton scattering
the related $N\bar{N}$ pair creation-process is  virtual,
in dispersion theory we have to treat the process described in
(\ref{t-channel})  in the unphysical region. 
In the cm frame of (\ref{t-channel}) where
\begin{equation} 
{\bf k_1}+{\bf k_2}=0 \label{k1plusk2}
\end{equation}
we obtain
\begin{equation}
\sqrt{t}=\sqrt{(k_1 + k_2)^2}= \omega_1 + \omega_2 = W^t,
\label{t-interpretation}
 \end{equation}
where  $W^t$ is the energy transferred to the $t$-channel via two-photon
fusion. At positive $t$ the $t$-channel
of Compton scattering $\gamma N\to N \gamma$ coincides with the $s$-channel
of the two-photon fusion reaction $\gamma_1+\gamma_2 \to N\bar{N}$.

\subsection{Hyperbolic dispersion relations}

Hyperbolic  dispersion
relations have been   introduced by Hite and Steiner \cite{hite73}
and further evaluated by
H\"ohler \cite{hohler83}  and Drechsel et al. \cite{drechsel03}. 
Hyperbolic dispersion relations  start with the ansatz
\begin{equation}
(s-a)(u-a)=(a-m^2)^2,\quad s+t+u=2m^2, \label{hy1}
\end{equation}
where a new parameter $a$ is introduced. From (\ref{hy1}) this new parameter
is derived in the form
\begin{equation}
a=\frac{(s-m^2)^2+st}{t},
\label{hy2}
\end{equation}
or in terms of the cm scattering angle $\theta_{\rm cm}$ 
as given in  (\ref{scatteringangle})
and the lab scattering angle
$\theta_{\rm lab}$:
\begin{equation}
a=-s\frac{1-\sin^2\frac12\theta_{\rm cm}}{\sin^2\frac12\theta_{\rm cm}},\quad
a=-m^2\frac{1-\sin^2\frac12\theta_{\rm lab}}{\sin^2\frac12\theta_{\rm lab}}
.
\label{hy3}
\end{equation}

After introducing the parameter $a$ hyperbolic dispersion relations
may be written down in the form
\begin{eqnarray}
&&{\rm Re} A_i(s,t,a)=A^B_i(s,t,a)+  A^{t-pole}_i(s,t,a)\nonumber\\
&&\quad\quad\quad\quad\quad\,\,\,\,+\frac{1}{\pi}{\cal P}
\int^{\infty}_{(m+m_\pi)^2}\,ds'\,{\rm Im}_sA_i(s',{\tilde
  t},a) \left[\frac{1}{s'-s}+\frac{1}{s'-u}-\frac{1}{s'-a}\right]\nonumber\\
&&\quad\quad\quad\quad\quad\,\,\,\,+\frac{1}{\pi}{\cal P}\int^\infty_0\,dt'\,
\frac{{\rm Im}_t A_i({\tilde s},t',a)}{t'-t} \label{hyperbolic}
\end{eqnarray}
where $A^B_i(s,t,a)$ is the Born term  and $A^{t-pole}_i(s,t,a)$
the pole in the $t$-channel.
The discontinuity in the $s$-channel, ${\rm Im}_s(s',{\tilde t},a)$, is
evaluated along the path given by
\begin{equation}
(s'-a)(u'-a)=(a-m^2)^2,\quad s'+{\tilde t}+u'=2m^2,
\label{path1}
\end{equation}
and  the corresponding quantity of the $t$-channel, ${\rm Im}_t\,A_i({\tilde
  s},t',a)$, along the path given by 
\begin{equation}
({\tilde s}-a)({\tilde u}-a)=(a-m^2)^2, \quad {\tilde s}+t'+{\tilde u}=2m^2.
\label{path2}
\end{equation}
The Mandelstam variables denoted by a tilde are those which do not
enter directly   into the dispersions integral, i.e. ${\tilde t}$
in case of the $s$-channel dispersion relation, and ${\tilde s}$ and 
${\tilde u}$ in case of the $t$-channel dispersion relation.

\subsection{Dispersion relations for the $s$-channel}

\begin{figure}[h]
\begin{center}
\includegraphics[width=0.8\linewidth]{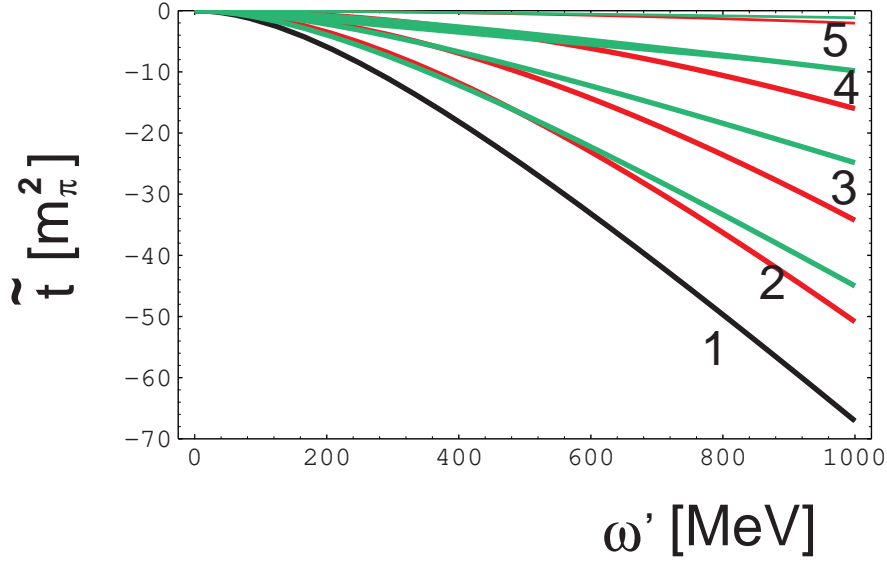}
\end{center}
\caption{Integration paths of hyperbolic dispersion relations for different 
quantities $a$: (1) $a=0$ ($\theta_{\rm cm}\equiv\theta_{\rm lab}=180^\circ$), 
(2) red or black curve: $a=-m^2$
  ($\theta_{\rm 
  lab}=90^\circ$), green or grey curve: $\theta_{\rm cm}=110^\circ$ , 
(3) red or black
curve: 
$a=-3\,m^2$ ($\theta_{\rm lab}= 60^\circ$), green or grey  curve: 
$\theta_{\rm cm}=75^\circ$
(4) red or black curve: $a=-10\,m^2$ ($\theta_{\rm
  lab}=35.1^\circ$), green or grey curve: $\theta_{\rm cm}=45^\circ$ , 
(5) red or black curve: $a=-100\,m^2$
($\theta_{\rm lab}=11.4^\circ$), green or grey curve: 
$\theta_{\rm cm}=15^\circ$. In
  the forward direction ($\theta_{\rm cm}\equiv\theta_{\rm lab}=0^\circ$) 
we have $a\to\infty$.}
\label{s-t-dia}
\end{figure}

Hyperbolic dispersion relations for the $s$-channel are given by the 
second line of 
Eq. (\ref{hyperbolic}). The last term in the square bracket is constructed
such that it provides integration paths located in the physical range
for all scattering angles. This is illustrated in Figure \ref{s-t-dia}
where integration paths are given in a ${\tilde t}-\omega'$ plane with
\begin{equation}
s'=m^2+2m\,\omega',
\label{so}
\end{equation}
relating the Mandelstam variable $s'$ to the photon energy $\omega'$
in lab frame.
The integration path (1) corresponds to $a=0$ and 
$\theta_{\rm lab}=180^\circ$, (2) red or black curve to $a=-m^2$ and 
$\theta_{\rm lab}=90^\circ$, (3) red or black curve to $a=-3\,m^3$ and 
$\theta_{\rm lab}=60^\circ$, (4) red or black curve to $a=-10\,m^2$ and 
$\theta_{\rm lab}=35.1^\circ$, (5) red or black curve to $a=-100\,m^2$ and 
$\theta_{\rm lab}= 11.4^\circ$. 
Hyperbolic dispersion relations correspond to integration paths 
with constant $\theta_{\rm lab}$. In principle this is equivalent to using
fixed-$\theta_{\rm cm}$ dispersion relations, except for the fact that slightly
different integration paths have to be applied in order  to cover  
the physical range 
of the ${\tilde t}-\omega'$-plane. This is shown in 
Figure \ref{s-t-dia} where also 
fixed-$\theta_{cm}$ integration paths are given, represented by the 
green or grey  curves. The only difference
 between these two options is that for the hyperbolic dispersion relations 
the $s$-channel parts have a somewhat more convenient form whereas the
$t$-channel parts are identical in the two cases.
It is interesting to note that the total backward hemisphere 
in the lab frame 
corresponds to
the small range between integration path (1) and (2), whereas the forward
hemisphere is spread out over a larger range in the physical
${\tilde t}-\omega'$ plane.
The important conclusion we may draw from this observation is that qualitative
properties of the scattering amplitudes  found at $\theta=180^\circ$ are also
valid at smaller angles inside the backward hemisphere. This means that
interference effects found at $\theta=180^\circ$ are also representative 
for smaller angles.

\subsection{Dispersion relations for the $t$-channel}

The $t$-channel singularities enter into the scattering amplitudes via
$t$-poles as given in the first line of Eq. (\ref{hyperbolic}). These
$t$-poles have an explicit $a$ or $\theta$ dependence except for
$\theta=180^\circ$ where the Eqs. (\ref{pseudoscalarPole-1}) 
and(\ref{scalarPole}) are
valid. 
Since explicit expressions for the $\theta$ dependence of the 
$t$-poles have not been derived up to now
this will be done in the following.

As an introduction we consider the well-known case of $\theta=180^\circ$
and then generalize the results by including smaller scattering angles.
At $\theta=180^\circ$ the kinematical constraints are 
\begin{equation}
s¸\,u=m^4,\quad\quad s+t+u=2\,m^2, \label{k1}
\end{equation}
being equivalent with
\begin{equation}
t=-\frac{(s-m^2)^2}{s}. \label{k2}
\end{equation} 
Using (\ref{k2}) we have to find the topological manyfold in the complex
$s$-plane where $t$ is a non-negative real number. This problem has been 
solved 
 in \cite{hearn62,koeberle68} where it has been shown that this topological
 manyfold 
is given by 
\begin{equation}
s=m^2\,e^{i\,\alpha} \label{k3}
\end{equation}
leading to 
\begin{equation}
t=4m^2\sin^2\frac{\alpha}{2}. \label{k4}
\end{equation}
This consideration shows that even in the case of $\theta=180^\circ$ where the
poles of
Eqs. (\ref{pseudoscalarPole-1}) and (\ref{scalarPole}) are independent of $s$  
the two variables $s$ and $t$ constrain each other.

It is important to note that the convenient representation (\ref{k3})
of the topological manifold is the only result borrowed from 
\cite{hearn62,koeberle68}. Especially we use  different dispersion 
relations and do not make explicit use of the circle (\ref{k3})
in the complex $s$-plane.
For the following only Eq. (\ref{k4}) is of importance which contains the
constraints of the variable $t$ in a transparent  way.

After this introductory remark it is easy to find the corresponding 
constraints for the
case of $\theta<180^\circ$. First we notice that the condition
\begin{equation}
(s-a)(u-a)=(a-m^2)^2\quad\quad s+t+u+=2m^2
\label{k5}
 \end{equation}
is equivalent with
\begin{equation}
x=\frac{t}{\sin^2\frac{\theta_{\rm cm}}{2}}=-\frac{(s-m^2)^2}{s}=
4m^2\sin^2\frac{\alpha}{2}. 
\label{k6}
\end{equation}
This means that for $\theta<180^\circ$ we have exactly the same kinematical 
relation as in case of $\theta=180^\circ$, except for the fact that the 
variable $t$ is replaced by the variable
\begin{equation}
x=\frac{t}{\sin^2 \frac{\theta_{\rm cm}}{2}}.
\label{k7}
\end{equation}

The $t$-channel part of the dispersion relation in (\ref{hyperbolic}) may 
now be written in terms of the parameter $x$ instead of the parameter $t$.
The advantage of this parameter transformation is that it takes care of the
kinematical constraints of Eq. (\ref{k5}).
Since $t$ is a nonnegative real number  (\ref{k6}) implies a 
constraint in the form
\begin{equation}
0\leq \sin^2\left(\frac{\alpha}{2}\right)\leq1, \quad 
0\leq t \leq t_{max}=4\,m^2\sin^2\left(\frac{\theta_{\rm cm}}{2}\right).
\label{constraint}
\end{equation}
Then with
\begin{equation}
x=4\,m^2\sin^2\left(\frac{\alpha}{2}\right),
\quad x_0=\frac{\mu^2}{\sin^2\left(\frac{\theta_{\rm cm}}{2}\right)}
\label{constraint1}
\end{equation}
we arrive at
\begin{eqnarray}
A^{t-pole}_i(t,\theta)={\cal M}(M\to\gamma\gamma)\,g_{MNN}\int^{4\,m^2}_0
\frac{\delta(x-x_0)}{x'-x}dx'=
\frac{{\cal M}(M\to\gamma\gamma)g_{MNN}}
{t-\mu^2}\sin^2\left(\frac{\theta_{\rm cm}}{2}\right)\\
 (i=1,2)\nonumber
\label{constraint2}
\end{eqnarray}
for a meson $M$ with mass $\mu$. 
 The amplitude given in (\ref{constraint2})
is the solution of the $t$-channel  integral of Eq.(\ref{hyperbolic}).

Eq. (\ref{constraint}) implies  that there exists a further constraint related
to the mass of the meson, given by
\begin{equation}
\mu^2\leq4\,m^2\sin^2\left(\frac{\theta_{\rm cm}}{2}\right).
\label{constraine3}
\end{equation}
Because of this relation, the finite masses of the mesons imply that the 
amplitudes $A_i(t,\theta)$ (i=1,2)
vanish below a lower limit
$\theta_{\rm lim}$ of the cm scattering angle.
These lower limits are
$\theta_{\rm lim}(\pi^0)=8.2^\circ$, 
$\theta_{\rm lim}(\eta)=34.0^\circ$,
$\theta_{\rm lim}(\sigma)=41.6^\circ$,  
$\theta_{\rm lim}(\eta')=61.3^\circ$ and 
$\theta_{\rm lim}(f_0,a_0)= 62.9^\circ$.

\subsection{Interference between nucleon and constituent-quark structure
  Compton scattering}

Hyperbolic dispersion relations as outlined above may be applied to a
calculation  of Compton differential cross sections in the whole angular range
 and for photon energies up to 1 GeV where precise photo-absorption cross
 sections are available. In this way the successful calculations 
\cite{lvov97,galler01,wolf01} in terms of
fixed-$t$ dispersions relations could be reproduced by  an independent
computational  procedure.

The most interesting application of course is the study of interference
phenomena of constituent-quark structure ($t$-channel)
and nucleon-structure ($s$-channel) Compton
scattering amplitudes. This will be carried out in the following
by investigating the energy dependence of the polarizabilities
$(\alpha-\beta)$ and $\gamma_\pi$.  
\begin{figure}[h]
\includegraphics[width=0.4\linewidth]{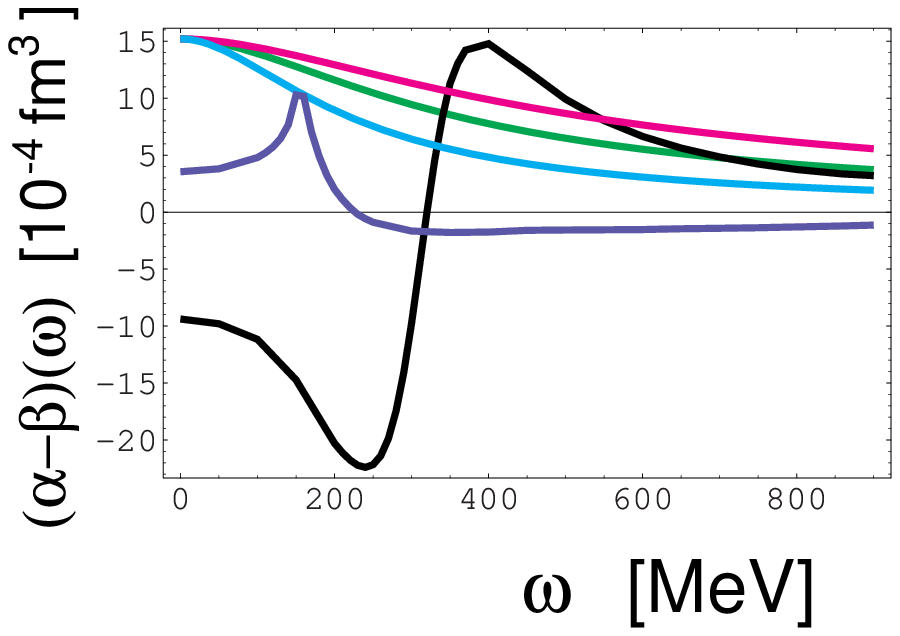}
\includegraphics[width=0.5\linewidth]{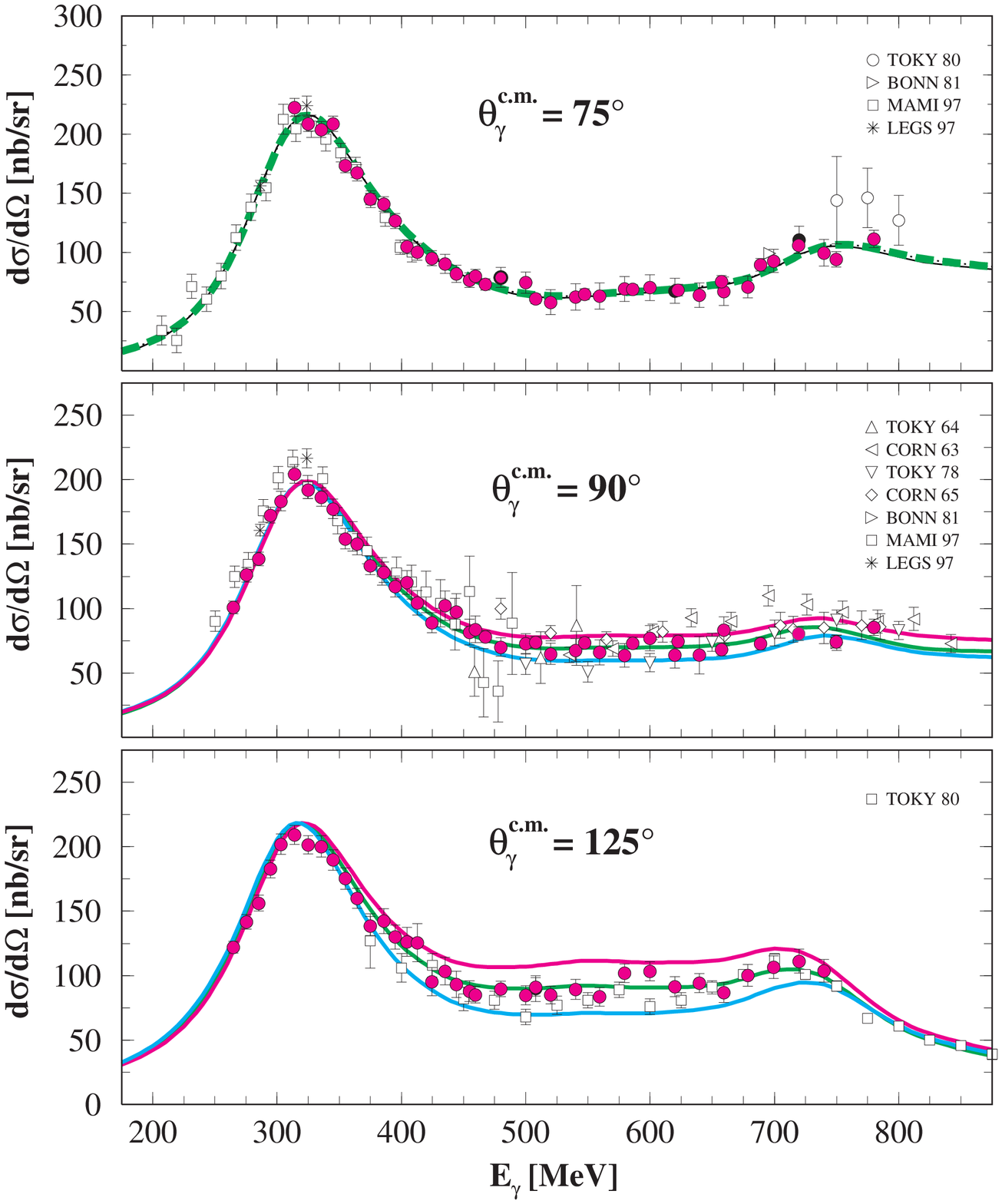}
\caption{Left panel: Difference of generalized polarizabilities 
$(\alpha-\beta)(\omega)$
versus photon energy $\omega$. Solid curve starting at
$(\alpha-\beta)(0)=-9.4$:
contribution of the $P_{33}(1232)$ resonance. Solid curve starting at
$(\alpha-\beta)(0)=+3.5$: Contribution of the nonresonant $E_{0+}$ amplitude.
Curves starting at $(\alpha-\beta)(0)=+15.2$: $t$-channel contribution of the
$\sigma$ meson calculated for different $\sigma$-meson masses. Upper curve
(dark grey or red): $m_\sigma=800$ MeV. Middle curve (grey or green):
$m_\sigma=600$ MeV. Lower curve (light grey or blue): $m_\sigma= 400$ MeV.
Not shown is the contribution of the $D_{13}(1520)$ resonance which cancels
the $E_{0+}$ contribution in the relevant energy range from 400 to 700 MeV.
Right panel: Differential cross sections for Compton scattering by the proton
versus photon energy. The three panels contain data corresponding to the
cm-angles of $75^\circ$, $90^\circ$ and $125^\circ$. The three curves 
are calculated for different mass parameters $m_\sigma=800$ MeV (upper, dark
grey or red), 600 MeV (center, grey or green) and 400 MeV (lower, light grey
or blue).\label{f1}
}
\end{figure}
\begin{figure}[h]
\centering\includegraphics[width=0.7\linewidth]{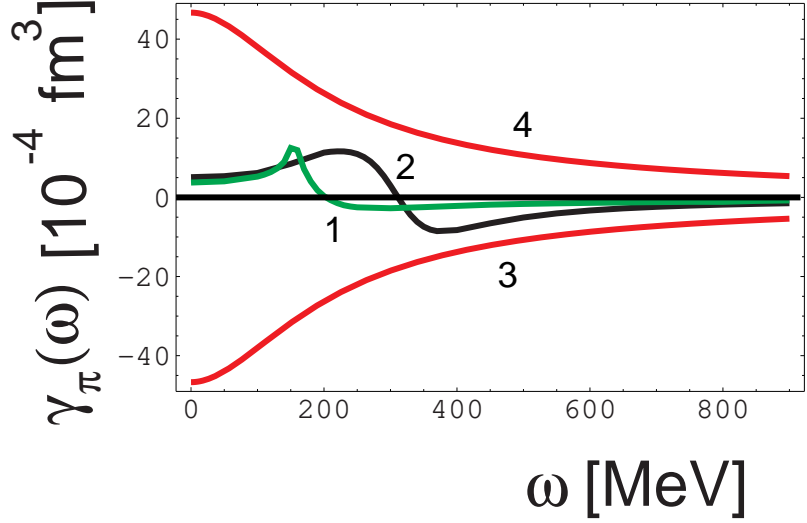}
\caption{Energy dependent backward spinpolarizability $\gamma_\pi(\omega)$
versus photon energy $\omega$ in the lab frame. Curve 1: Contribution due to 
the $E_{0+}$ CGLN amplitude, curve 2: Contribution due to the $P_{33}(1232)$
resonance, curve 3: $t$-channel contribution for the proton, curve 4:
$t$-channel contribution for the neutron.} 
\label{f2}
\end{figure}
For this purpose we introduce  linear combinations of invariant amplitudes
 which contain these  energy dependent 
polarizabilities \cite{schumacher05}:
\begin{eqnarray}
&&{\tilde A}_1(s,u,t)\equiv A_1(s,u,t)-\frac{t}{4m^2}A_5(s,u,t), \label{p1}\\
&&{\tilde A}_2(s,u,t)\equiv A_2(s,u,t)+\left(1-\frac{t}{4m^2}\right)
A_5(s,u,t). \label{p2}
\end{eqnarray}
Then the $\omega$-dependent ``dynamical''  polarizabilities
are given by
\begin{eqnarray}
&&
(\alpha-\beta)(\omega)=(\alpha-\beta)^s(\omega)+(\alpha-\beta)^t(\omega)
=-\frac{1}{2\pi}{\rm Re}{\tilde A}^{\rm nB}_1(s,t,a=0), \label{p1}\\
&&
\gamma_\pi(\omega)=\gamma^s_\pi(\omega)+\gamma^t_\pi(\omega)=-\frac{1}{2\pi\,m}
{\rm Re}{\tilde A}^{\rm nB}_2(s,t,a=0). \label{p2}
\end{eqnarray}
In (\ref{p1}) and (\ref{p2}) nB denotes that non-Born parts of the respective 
amplitudes
have to be taken into account. The $\omega$-dependent ``dynamical'' 
polarizabilities are
constructed such that 
they coincide with the conventional $\omega$-independent polarizabilities
in the limits $s=m^2$, $u=m^2$, and $t=0$.

The application of hyperbolic dispersion relations for $\theta=\pi$ 
leads to
\begin{eqnarray}
&&(\alpha-\beta)^s(\omega)=\frac{1}{2\pi^2}{\cal P}\int^\infty_{\omega_0}
\sqrt{1+2\frac{\omega'}{m}}F(\omega',\omega)
(\sigma(\omega',\Delta P={\rm yes})-\sigma(\omega',\Delta P={\rm no}))
\frac{d\omega'}{\omega'^2}, \nonumber\\
&&(\alpha-\beta)^t=-\frac{1}{2\pi}\frac{{\cal
      M}(\sigma\to\gamma\gamma) g_{\sigma NN}}{t(\omega)-m^2_\sigma}, 
\label{e1}
\end{eqnarray}
\begin{eqnarray}
&&\gamma^s_\pi(\omega)=\frac{1}{4\pi^2}{\cal P}\int^\infty_{\omega_0}
\sqrt{1+2\frac{\omega'}{m}}\left(1+\frac{\omega'}{m}\right)F(\omega',\omega)
\Sigma_n P_n\left(\sigma^n_{3/2}(\omega')-\sigma^n_{1/2}(\omega')\right)
\frac{d\omega'}{\omega'^3}, \nonumber\\
&&\gamma^t_\pi=-\frac{1}{2\pi\,m}\frac{{\cal
      M}(\pi^0\to\gamma\gamma) g_{\pi NN}}{t(\omega)-m^2_{\pi^0}}\tau_3, 
\label{e2}
\end{eqnarray}
with
\begin{eqnarray}
&&F(\omega',\omega)=\frac{\omega'^3+\omega'^2(m/2)}{
\omega'^3+\omega'^2 a(\omega)+\omega'b(\omega)+c(\omega)},\quad
\omega_0=m_\pi+\frac{m^2_\pi}{2m},\label{e3}\\
&& a(\omega)=\frac{-4\omega^2 +2m\omega+m^2}{2(m+2\omega)},\quad\quad\,\,
b(\omega)=-\frac{2m\omega^2}{m+2\omega},\label{e4}\\
&&  c(\omega) =-\frac{m^2\omega^2}{2(m+2\omega)},\quad\quad\quad \quad\quad\quad
t(\omega)=
-4\frac{\omega^2}{1+2\frac{\omega}{m}}. \label{e5}
\end{eqnarray}
The quantity $\Delta P$ denotes  parity change (yes) or nonchange (no), 
the parity factor is
$P_n=-1$ for $\Delta P=$ yes and $P_n=+1$ 
for $\Delta P=$ no.
The function $F(\omega',\omega)$ contains   the $\omega$
dependence  of the $s$-channel parts of the  polarizabilities and
converges like $F(\omega',\omega)\to 1$ for $\omega \to 0$.

The results obtained from the evaluation of Eq. (\ref{e1}) are shown in the
left panel of Figure \ref{f1}.  It is clearly demonstrated that there is a
constructive interference in the energy range from 400 to 700 MeV
between the contribution from the $P_{33}(1232)$
resonance and from the $t$-channel due to the $\sigma$ meson as part of the
constituent-quark structure. The sum of the two contributions depends on the
mass  
of the $\sigma$ meson assumed in the calculation and, therefore, may be used
for a determination of the $\sigma$-meson mass $m_\sigma$ from the differential
cross section for Compton scattering. This is outlined in the right panel of
Figure \ref{f1} where the effects of the $\sigma$-meson mass are most
prominent in the backward direction, as expected. The best agreement is
obtained for $m_\sigma=600$ MeV, being compatible with our present knowledge
of the ``bare'' mass$^1$ of the $\sigma$ meson. As pointed out before
 \cite{schumacher10} the experimental data shown in Figure \ref{f1} may be
 interpreted as a direct observation of the $\sigma$ meson while being part
 of the constituent-quark structure. Furthermore, this experiment provides us
 with the 
first and only direct determination of the ``bare'' mass$^1$ of 
the $\sigma$ meson,
whereas in all the other observations of the $\sigma$ meson 
the Breit-Wigner parameters or
the position of the
pole on the second
Riemann sheet are determined.

Eq.(\ref{p2})  provides a possibility to study
interference effects of the $t$-channel $\pi^0$ contribution with the
contribution from the $P_{33}(1232)$. Because of the smaller 
mass of the $\pi^0$
meson as compared with the $\sigma$ meson the relevant energy range 
is  at lower energies.  The interference pattern is
shown in Figure \ref{f2}. The largest interference effect is seen in the
energy range between 200 and 300 MeV. In this range the interference is
destructive in case of the proton and constructive 
in case of the neutron. 

In Figures \ref{f1} and \ref{f2} we have shown that there are pronounced
interference terms from the $t$-channel constributions and the 
contributions from the $P_{33}(1232)$ resonance for $(\alpha-\beta)(\omega)$
as well as for $\gamma_\pi(\omega)$. These interference terms in the two cases
allow to  determine an error for the mass $m_\sigma$ of the $\sigma$ meson 
due to uncertainties of the CGLN amplitude representing the $P_{33}(1232)$
resonance. The evaluation of $\gamma_\pi^{(p)}$ and $m_\sigma$ 
from the data  of the LARA experiment \cite{galler01,wolf01}
were based 
on the SAID-SM99K  parameterization,  with  the MAID2000
parameterization  used for comparison. The  results given in Table
\ref{LARA} were obtained
(see also \cite{schumacher05}):
\begin{table}[h]
\caption{Evaluation of $\gamma^{(p)}_\pi$ and $m_\sigma$ using data from the 
LARA experiment \cite{galler01,wolf01}}\vspace{3mm}\begin{center}
\begin{tabular}{llll}
\hline
&SAID-SM99K&MAID2000&\\
\hline
$\gamma^{(p)}_\pi$& $-37.1$&$-40.9$& LARA experiment \cite{galler01,wolf01}\\
$m_\sigma$& 600 MeV & 400 MeV& LARA experiment \cite{galler01,wolf01}\\
\hline
\end{tabular}\end{center}
\label{LARA}
\end{table}

\noindent
where LARA denotes the experiment leading to the results shown in Figure
\ref{f1}. For $\gamma^{(p)}_\pi$ the  SAID-SM99K parameterization leads to 
agreement with the present {\it recommended} value 
$\gamma^{(p)}_\pi=-36.4 \pm 1.5$ (experiment), -36.6 (predicted)
whereas MAID2000 leads to
results shifted by 10\%. 
By the same procedure $m_\sigma$ is shifted by 
$\Delta m_\sigma=- 200$ MeV.
Since the present {\it recommended} values are
based on the most recent and very precise CGLN amplitudes
\cite{drechsel07}
we can say that the evaluations of $\gamma^{(p)}_\pi$ and $m_\sigma$ 
on the bases  of the  SAID-SM99K parameterization
have been confirmed on a high level of precision, whereas  the use of 
the MAID2000 parameterization has been ruled out. Nevertheless, the shifts
of the results for $\gamma^{(p)}_\pi$ and $m_\sigma$ can be used for an error 
analysis for $m_\sigma$. For $\gamma^{(p)}_\pi$ the error amounts to about 1/3
of the shift and the same should be true for $m_\sigma$.
This consideration leads to an error of 
$\Delta m_\sigma = \pm  70$ MeV. Our final result therefore is
$m_\sigma= 600 \pm 70$ MeV.

\section{Summary and discussion}

In the foregoing paper we have outlined the present status of nucleon Compton
scattering and polarizabilities with emphasis on 
our present knowledge of 
spontaneous and dynamical chiral symmetry breaking in relation to
Compton scattering by the nucleon. An update of the list of experimental
{\it recommended} values for the polarizabilities is given. New results 
based on the experimental value for $\alpha_p$ and predicted values for
$(\alpha_n-\alpha_p)$ 
are presented which are strongly supporting  the {\it recommended}
experimental value for the electric
polarizability of the neutron. Especially, it appears extremely unlikely
that in future high-precisions experiments  a neutron electric
polarizability will be measured which is smaller than the corresponding
quantity for the proton.
New results
have been derived for the kinematical constraints of the $t$-channel
pole contributions and for hyperbolic dispersion relations. These results have
been used to study interference effects of scattering amplitudes due to the 
pion  cloud and the $P_{33}(1232)$ resonance ($s$-channel)
with scattering amplitudes due to
mesonic components of the structure of the constituent quarks ($t$-channel).
It has been shown
that scalar and pseudoscalar mesons as part of the constituent-quark
structure are
essential for a quantitative prediction of the polarizabilities $\alpha$,
$\beta$ and the spinpolarizability $\gamma_\pi$. Furthermore, effects of the 
$\sigma$ meson while being part of the constituent-quark structure
are clearly visible
in the experimental differential cross-section for Compton scattering 
by the proton. The analysis of experimental  differential
cross-sections in the second resonance region of the proton
shows that the ``bare'' mass$^1$ of the $\sigma$ meson
is $m_\sigma=600\pm 70$ MeV, being in agreement  with the 
QLL$\sigma$M  prediction 664 MeV within the margin of the error.

\end{document}